\documentclass[journal,twoside]{IEEEtran}
\ifCLASSINFOpdf
  \usepackage[pdftex]{graphicx}
\else
\fi
\usepackage{amsmath,amssymb,amsfonts}
\usepackage{mathtools}
\usepackage{graphicx}
\usepackage{subcaption}
\DeclareCaptionLabelFormat{ieee}{\textsc{#1}~#2}
\usepackage{textcomp}
\usepackage{xcolor}
\usepackage{tabularx}
\usepackage{booktabs}
\usepackage{float}
\usepackage{algpseudocode}
\usepackage[english]{babel}
\usepackage{paralist}
\usepackage{array}
\usepackage{amsmath,stackrel}
\usepackage{multicol}
\usepackage{verbatim}
\usepackage{enumerate}              
\usepackage{dsfont}
\usepackage{psfrag}
\usepackage[nolist]{acronym}
\usepackage[utf8]{inputenc}
\usepackage{tikz}
\usepackage{collcell}
\usepackage{pgfplots}
\pgfplotsset{compat=1.18}
\usepackage{hhline}
\usepackage{colortbl}
\usepackage{multido}
\usepackage{multirow}
\usepackage{dblfloatfix}    
\usepackage{balance}
\usepackage{etoolbox}
\usepackage{soul}
\usepackage{eucal}
\usepackage{cuted}
\usepackage[font=footnotesize]{caption}
\usepackage{relsize}

\usepackage{algorithm}
\usepackage{algpseudocode}
\usepackage{setspace}

\let\Algorithm\algorithm
\renewcommand\algorithm[1][]{\Algorithm[#1]\setstretch{1.1}}

\newcommand{\transp}{\mathrm{T}}

\usepackage{amsmath}
\usepackage{amssymb}
\usepackage{amsthm}
\usepackage{lipsum}
\usepackage{cuted}
\usepackage[bottom]{footmisc}

\usepackage{tabularx}
\usepackage{booktabs}
\usepackage[caption=false]{subfig} %
\usepackage{caption}
\usepackage{subcaption}

\makeatletter
\let\oldfootnote\footnote
\def\footnote{\@ifstar\footnote@star\footnote@nostar}
\def\footnote@star#1{{\let\thefootnote\relax\footnotetext{#1}}}
\def\footnote@nostar{\oldfootnote}
\makeatother
\usepackage{array}

\usepackage[nolist]{acronym}
\begin{acronym}
\acro{AdaGrad}{adaptive gradient}
\acro{AoA}{angle of arrival}
\acro{AoD}{angle of departure}
\acro{AWGN}{additive white Gaussian noise}
\acro{BS}{base station}
\acro{CCR}{correct classification rate}
\acro{CFAR}{constant false alarm rate}
\acro{CP}{cyclic prefix}
\acro{CRLB}{Cramér-Rao lower bound}
\acro{DBSCAN}{density-based spatial clustering of applications with noise}
\acro{DFT}{discrete Fourier transform}
\acro{DFT-s-OFDM}{discrete Fourier transform spread orthogonal frequency division multiplexing}
\acro{DF}{depth of focus}
\acro{ET}{extended target}
\acro{ELAA}{extremely large aperture array}
\acro{ELAS}{extremely large antenna spacing}
\acro{FAR}{false alarm rate}
\acro{FF}{far-field}
\acro{FFT}{fast Fourier transform}
\acro{FIM}{Fisher information matrix}
\acro{FOV}{field of view}
\acro{GD}{gradient descent}
\acro{GLRT}{generalized likelihood ratio test}
\acro{GML}{generalized maximum likelihood}
\acro{GOSPA}{generalized optimal sub-pattern assignment}
\acro{HPMW}{half-power mainlobe width}
\acro{ICI}{inter-carrier interference}
\acro{ISAC}{integrated sensing and communication}
\acro{ISI}{inter-symbol interference}
\acro{LFM}{linear frequency modulation}
\acro{LS}{least squares}
\acro{LLR}{log-likelihood ratio}
\acro{LoS}{line of sight}
\acro{MCR}{misclassification rate}
\acro{MF}{matched filter}
\acro{MIMO}{multiple-input multiple-output}
\acro{ML}{maximum likelihood}
\acro{MMSE}{minimum mean square error}
\acro{mmWave}{millimeter wave}
\acro{MB DFT-s-OFDM}{multi-block discrete Fourier transform spread orthogonal frequency division multiplexing}
\acro{NF}{near-field}
\acro{OFDM}{orthogonal frequency-division multiplexing}
\acro{PAPR}{peak-to-average power ratio}
\acro{pdf}{probability density function}
\acro{PEB}{position error bound}
\acro{PF}{particle filter}
\acro{pmf}{probability mass function}
\acro{PSK}{phase shift keying}
\acro{QPSK}{quadrature phase shift keying}
\acro{RF}{radio-frequency}
\acro{RoI}{region of interest}
\acro{Rx}{receiver}
\acro{RCS}{radar cross-section}
\acro{RMSE}{root mean square error}
\acro{SI}{self-interference}
\acro{SIC}{successive interference cancellation}
\acro{SNR}{signal-to-noise ratio}
\acro{SIMO}{single-input multiple-output}
\acro{THz}{terahertz}
\acro{Tx}{transmitter}
\acro{UE}{user equipment}
\acro{ULA}{uniform linear array}
\acro{XL-MIMO}{extremely large-scale MIMO}
\acro{6G}{sixth generation}
\end{acronym}

\usepackage{xcolor}

\begin{document}
\title{Extremely Large Antenna Spacing Method for Enhanced Wideband Near-Field Sensing}
\author{Tommaso~Bacchielli,~\IEEEmembership{Graduate Student Member,~IEEE,}        Lorenzo~Pucci,~\IEEEmembership{Member,~IEEE,}
and~Andrea~Giorgetti,~\IEEEmembership{Senior~Member,~IEEE}
\thanks{This work was supported by the European Union under the Italian National Recovery and Resilience Plan (NRRP) of NextGenerationEU, partnership on ``Telecommunications of the Future'' (PE00000001 - program ``RESTART'').}
\thanks{T. Bacchielli and A. Giorgetti are with the Department of Electrical, Electronic, and Information Engineering ``Guglielmo Marconi'' (DEI), University of Bologna, and the National Laboratory of Wireless Communications (WiLab), CNIT, Italy (e-mail: \{tommaso.bacchielli2, andrea.giorgetti\}@unibo.it). L. Pucci is with the National Laboratory of Wireless Communications (WiLab), CNIT, Italy (e-mail: lorenzo.pucci@wilab.cnit.it) \textit{T. Bacchielli and L. Pucci are co-first authors.}}
}


\maketitle

\begin{abstract}
This paper proposes a monostatic wideband system for integrated sensing and communication (ISAC) at millimeter-wave frequencies, based on multiple-input multiple-output (MIMO) orthogonal frequency-division multiplexing (OFDM). The system operates in a hybrid near-/far-field regime. The transmitter (Tx) operates in the far field (FF) and uses low-complexity beam steering. The receiver (Rx), on the other hand, operates in a pervasive near field (NF), enabled by a very large effective array aperture. To enable a fully digital implementation, we introduce an extremely large antenna spacing (ELAS) design. This design attains the required aperture with only a few widely spaced antenna elements while avoiding grating lobes in the composite Tx–Rx response. We analytically characterize the NF range-angle response of this architecture and study the interplay between NF effects and waveform bandwidth. This leads to the definition of a super-resolution region, where NF propagation at the Rx dominates the achievable range resolution and surpasses the classical, bandwidth-limited resolution. As a case study, we consider an extended target modeled as a collection of scatterers and assess localization performance via maximum-likelihood estimation. Numerical results evaluated in terms of root mean square error (RMSE) and generalized optimal sub-pattern assignment (GOSPA) show that operating in NF conditions with the ELAS-based design yields significant gains compared to a conventional FF baseline at both the Tx and Rx.
\end{abstract}


\IEEEpeerreviewmaketitle

\acresetall

\section{Introduction} \label{sec:intro}

\IEEEPARstart{T}{he} rapid evolution of large-scale antenna technologies and high-frequency operation has fundamentally transformed the landscape of wireless sensing, especially in the context of \ac{ISAC} for next-generation wireless systems of \ac{6G} and beyond \cite{Liu23,CarBar:L26}. As antenna apertures become extremely large, comparable to or even exceeding the operational wavelength by several orders of magnitude, the assumption of planar wavefronts, long adopted in classical \ac{FF} models, no longer holds \cite{WangBjo25}. Instead, electromagnetic propagation enters the \ac{NF} regime, where signals impinging on the array exhibit spherical wavefronts and spatially varying amplitude and phase profiles across the antenna aperture \cite{CuiDai22,GioBacGioDarDec:J25}. This transition to \ac{NF} propagation offers new opportunities for high-precision sensing and localization \cite{Dai25,He24}. In particular, in this regime, the assumption that the passive target can be approximated as a point scatterer becomes insufficient, and the concept of \acp{ET} must be considered \cite{Cong24,Sambon25}. In \ac{NF} sensing, the curvature of the incident wavefront encodes valuable range-dependent spatial information that is otherwise lost under the \ac{FF} approximation. This enables the joint estimation of range and angle parameters from a single array observation, effectively introducing an additional spatial degree of freedom \cite{Kosasih25}. As a consequence, \ac{NF} systems can better distinguish multiple target scatterers located at similar angular positions but at different ranges \cite{BelTagMizTebSpa:C25}. This capability results in an enhanced range resolution compared to \ac{FF} systems, where range resolution is solely determined by the bandwidth \cite{Durr21}. This property is especially beneficial for \ac{ET} localization, where the target is modeled as a collection of multiple scattering contributions with distinct spatial positions and reflection coefficients.

Recent studies have analyzed the impact of \ac{NF} on the sensing performance in both narrowband and wideband systems. In \cite{Bjo21}, the range resolution (or \ac{DF}) of an antenna under \ac{NF} conditions is derived as the distance interval where the array gain remains within 3 dB of its peak. The work in \cite{Kosasih24} explores how array shape and size influence the \ac{NF} beam behaviors in narrowband sensing systems and introduces a beam depth analysis for characterizing the \ac{NF} beam patterns in the distance domain. Similarly, \cite{WachPollin25} investigates the ambiguity function of narrowband \ac{NF} \ac{MIMO} radar for different aperture geometries. In wideband \ac{NF} systems, range resolution depends on the combined effect of bandwidth and \ac{NF}. The latter can enhance resolution when bandwidth is limited, reducing the need for large bandwidths. However, the \ac{NF} advantage on resolution diminishes with increasing target distance, unlike the robustness of wideband sensing, indicating that \ac{NF} sensing alone cannot fully replace the capabilities of wideband sensing \cite{WangMu24}. The study in \cite{Wach2025} approximates the range ambiguity function of wideband \ac{NF} systems, showing that when the resolution due to bandwidth is comparable to the \ac{NF} beam focusing, the \ac{NF} effect can improve the composite range profile by reducing sidelobes. Conversely, bandwidth expansion mitigates poor sidelobe levels in bandwidth-limited \ac{NF} systems. Works such as \cite{Rahal25,RahalTAES25} evaluate the performance of wideband \ac{OFDM}-based \ac{ISAC} systems that exploit both wavefront curvature and bandwidth to jointly estimate range and angle, enhancing spatial resolution. In \cite{WangWideband24} and \cite{WangTWC25}, authors analyze wideband \ac{NF} \ac{MIMO}-\ac{OFDM} systems, deriving \acp{CRLB} for multi-target localization and demonstrating that array aperture and bandwidth, rather than the number of antennas and subcarriers, primarily determine sensing accuracy.

Despite these advantages, the deployment of \acp{ELAA} introduces significant practical and architectural challenges. Achieving a large physical aperture by simply increasing the number of antenna elements, as in \ac{XL-MIMO} architectures, leads to prohibitively high hardware complexity and energy consumption \cite{Lu24}. In particular, implementing such arrays with a fully digital architecture—where each antenna element is equipped with its own \ac{RF} chain, including mixers, converters, and baseband processing units—is largely infeasible at large scale. In fact, this results in high power consumption and cost, and imposes stringent demands on synchronization, calibration, and real-time signal processing. These limitations have motivated the exploration of hybrid analog–digital architectures \cite{dehkordi2023multistatic,pucci24,Meng25,Luo25}, subarray structures \cite{Liu19,Zhu23}, and energy-efficient beamforming strategies to strike a balance between performance and implementation feasibility in large-scale \ac{NF} sensing systems with \acp{ELAA}.

To the best of the authors' knowledge, there remains a lack of research on practical, fully digital implementations of \acp{ELAA} that avoid the \ac{XL-MIMO} paradigm and its associated, unfeasibly large number of antenna elements. Furthermore, a thorough analysis of the benefits of wideband \ac{NF} sensing on range resolution and target position estimation accuracy, also supported by numerical simulations, has not yet been fully addressed. In particular, the literature does not clearly establish under which conditions the \ac{NF} effect at the \ac{Rx} can enhance the range resolution of a wideband sensing system beyond what is achievable by bandwidth alone.

When equipped with \acp{ELAA}, sensing systems can effectively exploit \ac{NF} propagation effects \cite{Qu24}. Such arrays are typically realized as \acp{ULA} with a very large physical aperture obtained by substantially increasing the number of antenna elements according to the \ac{XL-MIMO} paradigm \cite{Ye24}. The large aperture not only improves angular resolution but also extends the Fraunhofer distance, which defines the theoretical boundary between near- and \ac{FF} regions and grows quadratically with the aperture size \cite{Kosasih24}. As a result, \ac{NF} characteristics persist over distances that are relevant for sensing operation, allowing the system to achieve high spatial (range and angular) resolution and to better distinguish closely spaced scattering points. This ultimately enhances the reconstruction of \acp{ET} that can be modeled as a set of distributed scatterers \cite{dehkordi2023multistatic}.

To address these unexplored aspects, we take as a starting point the transmit–receive array configuration with dissimilar element spacings introduced in \cite{Fried12} for narrowband \ac{MIMO} radar, and generalize it to a fundamentally different operating regime—namely, a wideband \ac{MIMO} \ac{ISAC} system based on \ac{OFDM}, operating at \ac{mmWave} frequencies under hybrid \ac{NF} and \ac{FF} conditions.
%

This generalization departs from the narrowband, angle-only beampattern perspective of \cite{Fried12} by explicitly accounting for wideband signaling and near-field effects at the \ac{Rx}, enabled through an extremely large effective aperture achieved via an \ac{ELAS} design. The resulting framework allows us to investigate fundamental resolution limits and performance gains in hybrid near-/far-field \ac{ISAC} systems. In particular, the main contributions are summarized as follows:

\begin{enumerate}
    \item We design a monostatic wideband \ac{MIMO} \ac{OFDM}-based \ac{ISAC} architecture that, by design, drives the \ac{Rx} into a \ac{NF} operating regime. Specifically, the \ac{Tx} operates in the \ac{FF} and can therefore rely on low-complexity beam steering rather than beam focusing, while the \ac{Rx} operates under an all-pervasive \ac{NF} condition induced by the extremely large effective aperture enabled by the \ac{ELAS} design. This approach achieves \ac{NF} sensing capabilities with only a limited number of antenna elements, in contrast to the very large arrays typically required in \ac{XL-MIMO} systems.

    \item We analytically characterize the impact of waveform bandwidth on the proposed hybrid near-/far-field architecture, going beyond the narrowband, angle-only beampattern analysis in \cite{Fried12}. By studying the \ac{NF} range--angle response at the \ac{Rx}, we identify a super-resolution region around the transceiver in which \ac{NF} effects dominate the achievable range resolution and can significantly outperform classical bandwidth-limited resolution.

    \item We validate the proposed \ac{ELAS}-based design through numerical simulations, using a \ac{GLRT}-based framework for joint detection and target parameter estimation when modeling an \ac{ET} as a collection of scattering points. The sensing performance is assessed in terms of \ac{RMSE} and \ac{GOSPA} metrics, demonstrating clear performance gains when operating under \ac{NF} conditions at the \ac{Rx} compared to a conventional baseline assuming \ac{FF} propagation at both the \ac{Tx} and the \ac{Rx}.
\end{enumerate}

In this paper, bold uppercase and lowercase letters denote matrices and vectors, respectively. ${\mathbf{I}}_n$ is the $n\times n$ identity matrix, and $\mathrm{diag}(\cdot)$ represents the diagonal matrix. The operators $(\cdot)^\ast$, $(\cdot)^\mathrm{T}$, and $(\cdot)^\mathrm{H}$ denote conjugation, transpose, and conjugate transpose, respectively. We use $\mathrm{card}(\cdot)$, $|\cdot|$, $\|\cdot\|$, and $\otimes$ for cardinality, modulus, Euclidean norm, and Kronecker product, respectively, and $\mathbb{E}[\cdot]$ for expectation. A zero-mean circularly symmetric complex Gaussian random vector with covariance matrix $\boldsymbol{\Sigma}$ is denoted as $\mathbf{x} \sim \mathcal{CN}(\mathbf{0},\boldsymbol{\Sigma})$, and $x \sim {\mathcal{B}}(q,n)$ represents a Binomial random variable with number of trials $n$ and success probability $q$.

The remainder of the paper is organized as follows.
Section~\ref{sec:system_model} introduces the system model. Section~\ref{sec:ELAS} presents the proposed \ac{ELAS} method, while Section~\ref{sec:NF_B_r_res} analyzes the impact of near-field propagation and bandwidth on range resolution. Section~\ref{sec:estimation} describes the \ac{GLRT}-based detection and parameter estimation framework. Numerical results are reported in Section~\ref{sec:numerical_results}, and Section~\ref{sec:conclusions} concludes the paper.

\section{System Model} \label{sec:system_model}

\begin{figure}[t]
    \centering
    \includegraphics[width=0.9\columnwidth]{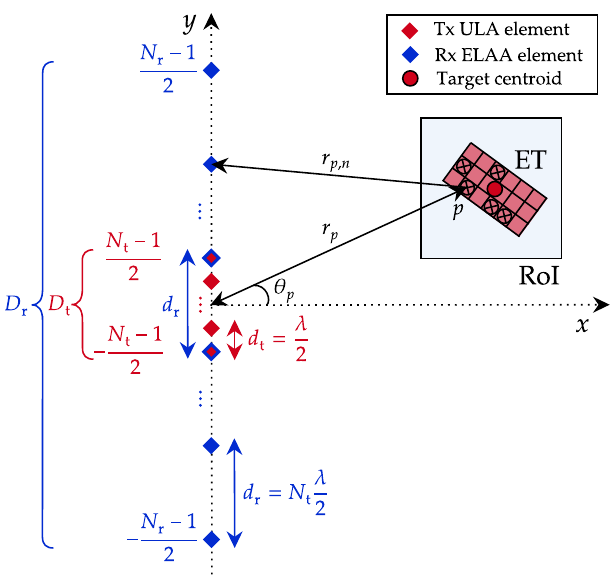}
    \caption{Monostatic MIMO ISAC setup with hybrid near-/far-field operation. The system is composed of a Tx ULA with half-wavelength spacing and a sparsely spaced Rx ELAA implementing the ELAS design to observe an ET located in the area of interest.}
    \label{fig:scenario}
\end{figure}

In this work, we consider a monostatic \ac{MIMO} \ac{OFDM}-based \ac{ISAC} transceiver with full-duplex operation, as depicted in Fig.~\ref{fig:scenario}.\footnote{A monostatic radar setup requires a full-duplex architecture employing analog/digital \ac{SI} cancellation methods (see, e.g., \cite{Barneto}). In this work, we assume that the \ac{SI} is effectively suppressed and therefore negligible compared to Gaussian noise at the sensing \ac{Rx}.} 

To perform sensing tasks, \ac{OFDM} frames consisting of $M$ \ac{OFDM} symbols with $K$ contiguous active subcarriers are transmitted. The $k$-th subcarrier has frequency $f_k = f_\mathrm{c} + (k - K/2)\Delta f$ with $\quad k = 0,\dots,K-1$, where $f_\mathrm{c}$ is the carrier frequency and $\Delta f$ is the subcarrier spacing. The total frame duration is $M T_\mathrm{s}$, where $T_\mathrm{s} \triangleq \frac{1}{\Delta f} + T_\mathrm{cp}$ is the \ac{OFDM} symbol duration including the \ac{CP}, and the total bandwidth is $B = K \Delta f$. Throughout this work, we assume that the system operates with $f_\mathrm{c} \gg B$, so that the narrowband array response assumption holds for each subcarrier \cite{VanTrees}.

Both the \ac{Tx} and \ac{Rx} are equipped with a \ac{ULA} aligned along the $y$-axis, composed of $N_\mathrm{t}$ and $N_\mathrm{r}$ antenna elements with inter-element spacings $d_\mathrm{t}$ and $d_\mathrm{r}$, respectively, with $d_\mathrm{r} \gg d_\mathrm{t}$ \cite{Fried12}. The \ac{Tx} array is placed at $\mathbf{p}^\mathrm{tx} = [0,\,0]^\mathrm{T}$, and has an aperture $D_\mathrm{t} = (N_\mathrm{t}-1)d_\mathrm{t}$. We consider relatively small transmit apertures such that the corresponding Fraunhofer distance $D_\mathrm{F}^\mathrm{tx} = 2 D_\mathrm{t}^2/\lambda$ is much smaller than the distances of interest in the monitored area. Hence, the \ac{Tx} can be modeled as operating under \ac{FF} propagation conditions.

The $n$-th \ac{Rx} antenna element is located at
\begin{equation}
    \mathbf{p}^\mathrm{rx}_n = [x^\mathrm{rx}_n,\,y^\mathrm{rx}_n]^\mathrm{T}
    = \left[0,\,\tilde{n}d_\mathrm{r}\right]^\mathrm{T}, \qquad n = 1,\dots,N_\mathrm{r}
    \label{eq:p_n}
\end{equation}
with $\tilde{n} = \frac{2n -N_\mathrm{r}-1}{2}$, leading to the \ac{Rx} aperture
\begin{equation}
    D_\mathrm{r} = (N_\mathrm{r}-1)d_\mathrm{r}.
\end{equation}
We assume that the sensing system monitors an area whose maximum Euclidean distance from the transceiver, denoted by $r_\mathrm{max}$, satisfies
\begin{equation}
    r_\mathrm{max} \ll D_\mathrm{F}^\mathrm{rx} = \frac{2 D_\mathrm{r}^2}{\lambda}
    \label{eq:r_max}
\end{equation}
where $D_\mathrm{F}^\mathrm{rx}$ denotes the Fraunhofer distance with respect to the \ac{Rx}, $\lambda = c / f_\mathrm{c}$ is the wavelength and $c$ is the speed of light. Under this condition, the propagation with respect to the \ac{Rx} is such that \ac{NF} effects can be exploited for sensing.

The specific choice of the inter-element spacings $d_\mathrm{t}$ and $d_\mathrm{r}$ is based on an antenna spacing design that enables a huge effective aperture at the \ac{Rx} while using a limited number of antenna elements. This design, referred to as the \ac{ELAS} method, is inspired by the approach proposed in \cite{Fried12} and is detailed in Section~\ref{sec:ELAS}.

\subsection{Input-Output Relationship} \label{sec:I/0}

The complex baseband \ac{OFDM} signal transmitted by the \ac{Tx} antenna array is expressed in vector form as\footnote{To simplify the notation, the \ac{CP} is omitted in \eqref{eq:tx_signal}, although it is implicitly included to guarantee robustness against \ac{ISI}.}
\begin{equation}
    \mathbf{s}(t) = \sqrt{\frac{P_\mathrm{t}G_\mathrm{t}}{K}}\mathbf{w}_\mathrm{t} \sum_{m=0}^{M-1}\left( \sum_{k=0}^{K-1}x[k,m]e^{\jmath2 \pi k \Delta f t}\right)g_\mathrm{tx}(t-m T_\mathrm{s})
    \label{eq:tx_signal}
\end{equation}
where $P_\mathrm{t}$ is the total transmit power, $G_\mathrm{t}$ is the single antenna element gain at the \ac{Tx}, $\mathbf{w}_\mathrm{t}\in \mathbb{C}^{N_\mathrm{t} \times 1}$ is the beamforming weight vector designed based on the application (more details will be provided later), $x[k,m]$ denotes a generic complex modulation symbol transmitted at time instant $m$ on subcarrier $k$ and normalized so that $\mathbb{E}\bigl\{\bigl|x[k,m]\bigr|^2\bigr\} = 1$, while $g_\mathrm{tx}(t)$ represents the modulating pulse.

At the \ac{Rx}, after the \ac{FFT} block for \ac{OFDM} demodulation, a received time-frequency grid of complex elements $y[k,m]$ is obtained at each antenna element. By considering negligible \ac{ICI} and \ac{ISI}, the vector $\mathbf{y}[k,m] \in \mathbb{C}^{N_\mathrm{r} \times 1}$ of received complex modulation symbols at subcarrier $k$ for $m$-th \ac{OFDM} symbol is given by
\begin{equation} \label{eq:Rx_signal}
    \mathbf{y}[k,m] = \sqrt{\frac{P_\mathrm{t}G_\mathrm{t}G_\mathrm{r}}{K}}\mathbf{H}[k,m] \mathbf{w}_\mathrm{t} \, x[k,m] +\boldsymbol{\nu}[k,m]
\end{equation}
where $G_\mathrm{r}$ is the single antenna element gain of the \ac{Rx}, $\mathbf{H}_{\mathrm{t},\mathrm{r}}[k,m] \in \mathbb{C}^{N_\mathrm{r} \times N_\mathrm{t}}$ is the \ac{MIMO} channel matrix in the frequency domain 
for the subcarrier $k$ at time $m$ which is defined in Section~\ref{sec:channel_model}. Moreover, $\boldsymbol{\nu}[k,m] \sim \mathcal{CN}(\mathbf{0},\sigma_{\nu}^2\mathbf{I}_{N_\mathrm{r}})$ represents the \ac{AWGN}, with noise variance $\sigma_{\nu}^2= N_0 \Delta f$, where $N_0 = k_\mathrm{B} T_0 n_\mathrm{F}$ is the noise power spectral density, $k_\mathrm{B}$ is the Boltzmann constant, $T_0=290\,$K, and $n_\mathrm{F}$ is the \ac{Rx} noise figure.

\subsection{Channel and Target Model} \label{sec:channel_model}

Without loss of generality, this work considers a single \ac{ET} scenario, where the \ac{ET} is modeled as a set of grid elements $\mathcal{E}$ within a designated rectangular region $\mathcal{A} \subset \mathbb{R}^2$ with a fixed size of length $L$ and width $W$ and an area of $|\mathcal{A}|=LW$, as proposed in \cite{dehkordi2023multistatic} (see Fig.~\ref{fig:scenario}).
The decision to model the target as an extended object is based on the assumption that the \ac{Rx} is operating in \ac{NF} conditions. Therefore, it is reasonable to assume that the target can be seen as a set of scatterers from the \ac{Rx} perspective. 

At each time measurement, the \ac{ET} is composed of a random number $P<|\mathcal{E}|$ of micro-scatterers. Given that each grid element inside $\mathcal{A}$ can be active, i.e., the scatterer is localized inside that element, with probability $q$, the number of active points, or target scatterers, follows a binomial distribution, i.e., $P \sim \mathcal{B}(q,|\mathcal{E}|)$. With an appropriate choice of the grid size, the varying number of scatterers in the target area can effectively model the fluctuations and variance of an object’s radar reflectivity caused by factors such as, e.g., target aspect angle and material. Hereinafter, we refer to the scatterers of the \ac{ET} as micro-scatterers to emphasize that they represent reflective points of a larger object. 

Considering \ac{LoS} propagation conditions and accounting for all the visible micro-scatterers from the \ac{ET}, the $N_\mathrm{r} \times N_\mathrm{t}$ hybrid near-/far-field \ac{MIMO} channel matrix in \eqref{eq:Rx_signal} for subcarrier $k$ at time $m$ can be written as
\begin{equation} \label{eqn:H}
    \mathbf{H}[k,m] = \sum_{p=1}^{P}\epsilon_p e^{\jmath2\pi(mT_\mathrm{s}f_{\mathrm{D},p}-k\Delta f\tau_p)} \mathbf{b}(\theta_{p},r_p) \mathbf{a}^{\mathrm{H}}(\theta_{p})
\end{equation}
where $r_p$ is the reference Euclidean distance between the center of the transceiver and the $p$-th micro-scatterer, $\tau_p=2r_p/c$ is the corresponding reference round-trip delay related to the micro-scatterer $p$ and $f_{\mathrm{D},p}=2v_p f_\mathrm{c}/c$ is the reference Doppler shift associated with the $p$-th micro-scatterer, with $v_p$ its relative radial velocity. Moreover, $\epsilon_p=\sqrt{\frac{\sigma_p c^2}{(4\pi)^3f_\mathrm{c}^2r_p^4}}e^{-\jmath2\pi f_\mathrm{c}\tau_p}$ is the reference complex channel factor, and $\sigma_p$ is the \ac{RCS} associated with micro-scatterer $p$. The \ac{FF} transmit and \ac{NF} receive array responses, denoted by $\mathbf{a}(\theta_{p}) \in \mathbb{C}^{N_\mathrm{t} \times 1}$ and $\mathbf{b}(\theta_{p},r_p)~\in~\mathbb{C}^{N_\mathrm{r} \times 1}$, respectively, are defined as \cite{dehkordi2023multistatic}
\begin{align} \label{eq:steering_vec_Tx}
    \mathbf{a}(\theta_{p})=&\left[e^{-\jmath\frac{N_\mathrm{t}-1}{2}\pi \sin{(\theta_{p})}},\dots,e^{\jmath\frac{N_\mathrm{t}-1}{2}\pi \sin{(\theta_{p})}}\right]^\mathrm{T}\\
    \nonumber\\
    \mathbf{b}(\theta_{p},r_p)=&\Bigr[\frac{r_p}{r_{p,1}}e^{-\jmath\frac{2\pi}{\lambda}(r_{p,1}-r_p)}, \dots,\frac{r_p}{r_{p,n}}e^{-\jmath\frac{2\pi}{\lambda}(r_{p,n}-r_p)}, \nonumber\\
    & \quad\dots,\frac{r_p}{r_{p,{N_\mathrm{r}}}}e^{-\jmath\frac{2\pi}{\lambda}(r_{p,{N_\mathrm{r}}}-r_p)}\Bigl]^\mathrm{T}
    \label{eq:steering_vec_Rx}
\end{align}
where $r_p$ and $\theta_p$ are the reference distance and reference \ac{AoD}/\ac{AoA}, respectively, between the \ac{Tx}/\ac{Rx} and the scatterer $p$, calculated with respect to the geometric center of the array at $[0,0]^\mathrm{T}$.\footnote{Note that, in practice, the \ac{Tx} and \ac{Rx} antennas may be separated by a small distance. Here, for simplicity, we assume that their phase centers coincide at the origin. This approximation is accurate when the \ac{Tx}–\ac{Rx} separation is negligible compared to the target ranges, and any residual offset can be accounted for through calibration.} 
In \eqref{eq:steering_vec_Rx}, $r_{p,n}$ denotes the Euclidean distance between the scatterer $p$ and the $n$-th receive antenna. If the scatterer $p$ is located at
$\mathbf{p}_{p} = [x_{p},y_{p}]^\mathrm{T} = [r_p\cos(\theta_p),\,r_p\sin(\theta_p)]^\mathrm{T}$, this distance can be expressed as \cite{Liu23}
\begin{equation}
    r_{p,n} = \|\mathbf{p}_{p}-\mathbf{p}^\mathrm{rx}_n\|
    = r_p\sqrt{1+\frac{(\tilde{n} d_\mathrm{r})^2}{r_p^2}-\frac{2 \tilde{n} d_\mathrm{r}\sin(\theta_p)}{r_p}}.
\end{equation}
where $\tilde{n}$ is provided in \eqref{eq:p_n}.

A second-order Taylor expansion can accurately approximate such a distance, also known as the Fresnel approximation \cite{Tao11,Fried19,Gio25}, i.e.,
\begin{equation}
    r_{p,n} \approx r_p - \tilde{n} d_\mathrm{r}\sin(\theta_p)
    + \frac{(\tilde{n} d_\mathrm{r}\cos(\theta_p))^2}{2 r_p}.
\end{equation}
Note that, under \ac{FF} conditions, $r_p \simeq r_{p,n}$ and $r_p \gg \tilde{n} d_\mathrm{r}$ for all $n$. In this regime, the \ac{Rx} array response loses its dependence on the distance $r_p$ and reduces to the well-known \ac{FF} array response vector as per \eqref{eq:steering_vec_Tx}.

\section{Proposed Extremely Large Antenna Spacing (ELAS) Method for Near-Field Sensing} \label{sec:ELAS}

As previously mentioned, in this work, we deliberately consider a \ac{Tx} array with a limited aperture, so that its Fraunhofer distance $D_\mathrm{F}^\mathrm{tx}$ is much smaller than the target ranges of interest. Under this design choice, the \ac{Tx} can be modeled as operating in the \ac{FF} for all points in the monitored area, thus eliminating the need for a beam focusing scheme at the \ac{Tx}, which can be challenging and computationally demanding for some applications due to the joint search over range and angle \cite{dehkordi2023multistatic}. Instead, a straightforward beam steering approach is adopted to illuminate the monitored area. Conversely, the considered \ac{Rx} is an \ac{ELAA} with an aperture large enough for the entire monitored area to lie in the effective \ac{NF} of the antenna. \footnote{It is worth noting that the effective \ac{NF} region, from a sensing perspective, is typically much smaller than the region implied by the Fraunhofer distance, approximately within one seventh of that distance \cite{WachPollin25}. In fact, the classical Fraunhofer distance is defined based on the phase error of the far-field approximation with respect to the accurate \ac{NF} model, which may not effectively reflect sensing performance in terms of resolution and localization accuracy \cite{Liu23,WangBjo25}.} 
This configuration enables enhanced spatial (range and angle) resolution, thereby improving target localization accuracy.

A common way to realize such an \ac{ELAA} is to employ a \ac{ULA} with a very large number of antennas and an inter-element spacing of $\lambda/2$. This avoids undesired grating lobes (i.e., ambiguities) in the angle domain. However, such a design may be challenging from a hardware implementation viewpoint and highly expensive. In fact, when the number of antennas becomes very large, current fully digital and hybrid implementations do not scale well in terms of complexity, power consumption, and latency.
\begin{figure*}[t]
\centering
\begin{subfigure}[b]{0.32\textwidth}
    \centering
    \includegraphics[width=\columnwidth]{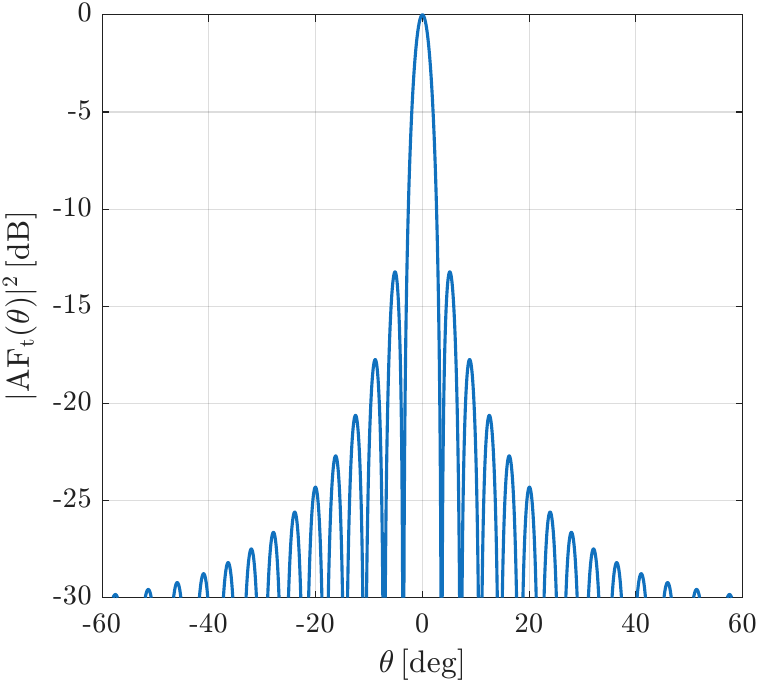}
    \caption{}
    \label{fig:AF_Tx}
\end{subfigure}
\hfill
\begin{subfigure}[b]{0.32\textwidth}
    \centering
    \includegraphics[width=\columnwidth]{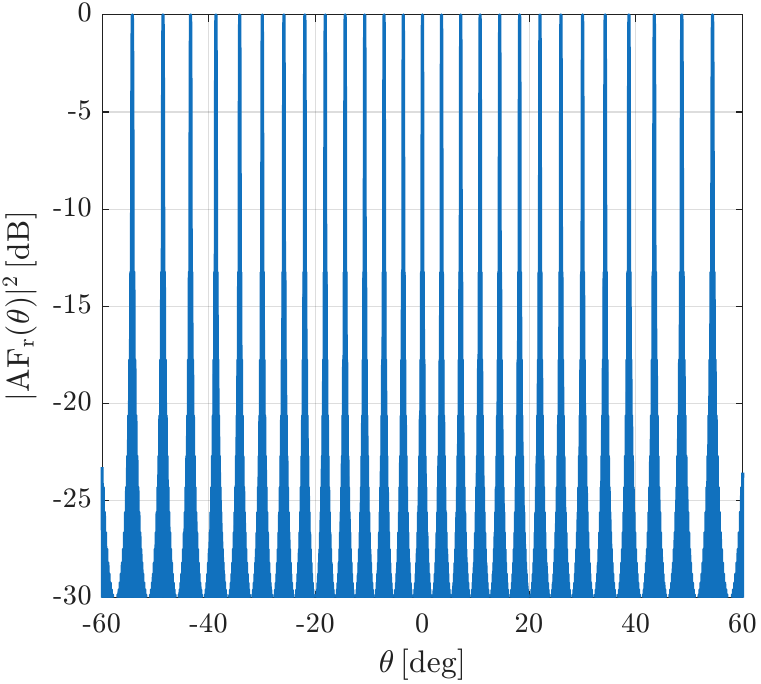}
    \caption{}
    \label{fig:AF_Rx_ang}
\end{subfigure}
\hfill
\begin{subfigure}[b]{0.32\textwidth}
    \centering
    \includegraphics[width=\columnwidth]{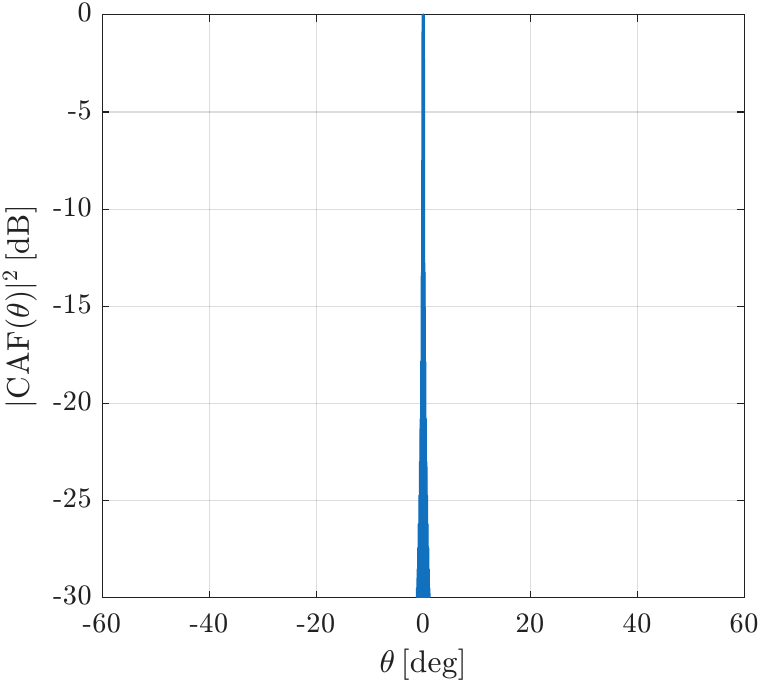}
    \caption{}
    \label{fig:CAF_ang}
\end{subfigure}
\caption{Normalized Tx FF array factor (a), Rx NF array factor (b), and composite array factor (c) evaluated in the angular domain under the angular sampling method assumption, by considering the proposed monostatic radar configuration with the following system parameters: $f_\mathrm{c}=60\,$GHz, $d_\mathrm{t}=\lambda/2$, $d_\mathrm{r}=N_\mathrm{t}\lambda/2$ and $N_\mathrm{t}=N_\mathrm{r}=32$.}
\label{fig:AF}
\end{figure*}

To explore a lower-complexity solution, we adopt an antenna spacing scheme where the \ac{Tx} and \ac{Rx} arrays are designed to yield a composite response equivalent to that of an $N_\mathrm{t}N_\mathrm{r}$-element array with $\lambda/2$ spacing, as in \cite{Fried12}. In particular, the \ac{Tx} antennas are spaced by $d_\mathrm{t} = \lambda/2$, whereas at the \ac{Rx} the inter-element spacing is chosen as $d_\mathrm{r}=N_\mathrm{t}\,\lambda/2$, i.e., much larger than $\lambda/2$. Such a spacing at the \ac{Rx} would, in principle, generate grating lobes (ambiguities) in the angle response. However, thanks to the combination of the \ac{Tx} and \ac{Rx} array responses, the grating lobes are effectively suppressed in the composite \ac{Tx}–\ac{Rx} array response. While the analysis in \cite{Fried12} is limited to the composite beampattern in a narrowband, \ac{FF} regime, in the following, we explicitly characterize the individual transmit and receive angle responses, as well as their composite response. We show that this architecture realizes a hybrid near-/far-field system, with the \ac{Tx} operating in the \ac{FF} (i.e., with reduced complexity) and the \ac{Rx} in the \ac{NF} (i.e., with enhanced target localization accuracy), and analyze the impact of the signal bandwidth on localization performance.

\subsection{Far-Field Tx Array Factor} \label{sec:AF_Tx}

As previously described, the \ac{Tx} employs a \ac{ULA} with $N_\mathrm{t}$ elements and spacing $d_\mathrm{t}=\lambda/2$, operating in the \ac{FF} over the monitored area. With reference to \eqref{eq:steering_vec_Tx}, we adopt a beam steering approach with beamforming weights $\mathbf{w}_\mathrm{t}=\mathbf{a}(\overline{\theta})$, where $\overline{\theta}$ is the desired pointing direction, where, e.g., a potential target can be present.

The normalized \ac{Tx} array factor is given by
\begin{align} \label{eqn:AF_Tx} \nonumber
    \mathrm{AF}_\mathrm{t}(\theta)&=\frac{1}{N_\mathrm{t}}\mathbf{a}^\mathrm{H}(\theta)\mathbf{w}_\mathrm{t} = \frac{1}{N_\mathrm{t}}\mathbf{a}^\mathrm{H}(\theta)\mathbf{a}(\overline{\theta})\\ \nonumber
    &=\frac{1}{N_\mathrm{t}} \sum_{n=-\frac{N_\mathrm{t}-1}{2}}^{\frac{N_\mathrm{t}-1}{2}} e^{\jmath\pi n\left(\sin{(\overline{\theta})}-\sin{(\theta)}\right)}\\
    &= \frac{1}{N_\mathrm{t}} \frac{\sin{\left(N_\mathrm{t}\pi/2\left(\sin{(\theta)}-\sin{(\overline{\theta})}\right)\right)}}{\sin{\left(\pi/2\left(\sin{(\theta)}-\sin{(\overline{\theta})}\right)\right)}}.
\end{align}

Note that, the expression of the normalized \ac{Tx} array factor $\mathrm{AF}_\mathrm{t}$ in \eqref{eqn:AF_Tx} has the form of the \emph{Dirichlet kernel} $\mathrm{D}_{\frac{N_\mathrm{t}-1}{2}}\left(\left(\sin{(\theta)}-\sin{(\overline{\theta})}\right)/2\right)$. From the properties of this periodic function, its amplitude exhibits a main lobe in correspondence with the illuminated direction  $\theta=\overline{\theta}$ and many secondary lobes of decreasing amplitude with periodic notches at
\begin{equation} \label{eqn:notches_Tx}
\theta_{\mathrm{notch},i}^\mathrm{tx}=\arcsin{\left((2i/N_\mathrm{t})+\sin{(\overline{\theta})}\right)}
\end{equation}
for $i \in \mathbb{Z} \setminus \{0\}$, as illustrated in Fig.~\ref{fig:AF_Tx}. The angular resolution of the transmit array is determined by the main-lobe width of its (normalized) power beampattern $|\mathrm{AF}_\mathrm{t}(\theta)|^2$, and is conventionally quantified by the $3$\,dB beamwidth, i.e., $\Delta\theta_{3\mathrm{dB}}^\mathrm{tx} \approx 0.89\lambda/D_\mathrm{t} \approx 2/N_\mathrm{t}$ \cite{Richards}.

\subsection{Near-Field Rx Array Factor} \label{sec:AF_Rx}

The \ac{Rx} is designed to operate as an \ac{ELAA}, providing effective \ac{NF} conditions over the entire monitored area. As already mentioned, this is achieved by adopting the \ac{ELAS} design \cite{Fried12}, in which the \ac{Rx} ULA has $N_\mathrm{r}$ elements with inter-element spacing $d_\mathrm{r} = N_\mathrm{t}\,\frac{\lambda}{2}$, thus realizing a very large aperture by increasing the spacing rather than the number of antennas. This choice contrasts with the conventional $\lambda/2$ spacing typically used to avoid grating lobes and ambiguity issues in \ac{AoA} estimation.

To derive the \ac{Rx} array factor under the above assumptions, we consider a beam focusing approach in which the receive weight vector $\mathbf{w}_\mathrm{r} \in \mathbb{C}^{N_\mathrm{r} \times 1}$ is given by $\mathbf{w}_\mathrm{r}=\mathbf{b}(\overline{\theta},\overline{r})$ with $\mathbf{b}(\theta,r)$ reported in \eqref{eq:steering_vec_Rx}. Here, $\left(\overline{\theta},\overline{r}\right)$ represents the polar coordinates of a generic focus point with respect to the center of the \ac{Rx} antenna array. The normalized \ac{NF} array factor at the \ac{Rx} can be expressed as
\begin{align} \label{eqn:AF_Rx} \nonumber
    \mathrm{AF}_\mathrm{r}(\theta,r)&=\frac{1}{N_\mathrm{r}} \mathbf{w}_\mathrm{r}^\mathrm{H}\mathbf{b}(\theta,r) = \frac{1}{N_\mathrm{r}} \mathbf{b}^\mathrm{H}(\overline{\theta},\overline{r})\mathbf{b}(\theta,r)\\ \nonumber
    &\approx \frac{1}{N_\mathrm{r}} \sum_{n=-\frac{N_\mathrm{r}-1}{2}}^{\frac{N_\mathrm{r}-1}{2}} e^{-\jmath\pi N_\mathrm{t}^2\frac{\lambda}{4} n^2 \left(\frac{\cos^2{(\theta)}}{r}-\frac{\cos^2{(\overline{\theta})}}{\overline{r}}\right)}\\
    &\times e^{\jmath\pi N_\mathrm{t}n\left(\sin{(\theta)}-\sin{(\overline{\theta})}\right)}.
\end{align}

Note that the \ac{NF} array factor at the \ac{Rx} in \eqref{eqn:AF_Rx} cannot, in general, be expressed with a simple closed form due to its joint dependence on the \ac{AoA} $\theta$ and the range $r$. To gain further insight, we follow the approach in \cite{CuiDai22}, which introduces two sampling methods, the \emph{angular sampling method} and the \emph{distance sampling method}, that decouple the array factor into angle-only and range-only components under suitable assumptions. In the following, we employ these methods to study the \ac{Rx} array factor separately in the angular and range domains and derive closed-form expressions in each domain.

\subsubsection{Near-field Rx array factor in the angular domain} \label{sec:AF_Rx_ang}

The derivation of the \ac{Rx} array factor in \eqref{eqn:AF_Rx} in the angular domain follows the angular sampling method in \cite{CuiDai22}. The key idea is to evaluate $\mathrm{AF}_\mathrm{r}$ at points in polar coordinates $(\theta,r)$ that satisfy
\begin{equation}
    \frac{\cos^2(\theta)}{r} = \frac{\cos^2(\overline{\theta})}{\overline{r}} = \frac{1}{R}
\end{equation}
where $R>0$ is a given constant. Equivalently, we restrict the analysis to locations lying on the curve $r = R \cos^2(\theta)$,  which we refer to as the \emph{distance ring} $R$, in analogy with \cite{CuiDai22}.

Under this condition, the quadratic phase term in \eqref{eqn:AF_Rx}, which originally depends jointly on angle and range, vanishes, and the normalized array factor becomes only a function of the \ac{AoA}, as follows
\begin{align} \label{eqn:AF_Rx_ang} \nonumber
    \mathrm{AF}_\mathrm{r}(\theta)&\approx\frac{1}{N_\mathrm{r}} \sum_{n=-\frac{N_\mathrm{r}-1}{2}}^{\frac{N_\mathrm{r}-1}{2}} e^{\jmath\pi N_\mathrm{t} n\left(\sin{(\theta)}-\sin{(\overline{\theta})}\right)}\\
    &= \frac{1}{N_\mathrm{r}} \frac{\sin{\left(N_\mathrm{r}N_\mathrm{t}\pi/2\left(\sin{(\theta)}-\sin{(\overline{\theta})}\right)\right)}}{\sin{\left(N_\mathrm{t}\pi/2\left(\sin{(\theta)}-\sin{(\overline{\theta})}\right)\right)}}.
\end{align}
It is worth noting that, in those regions of space where the range dependence can be removed, the \ac{NF} array factor at the \ac{Rx} observed in the angular domain exhibits the same Dirichlet-kernel structure as the \ac{FF} \ac{Tx} array factor in \eqref{eqn:AF_Tx}, but with an effectively larger aperture. However, differently from the \ac{Tx}, the \ac{Rx} array response shows periodic grating lobes at
\begin{equation} \label{eqn:gl_Rx}
    \theta_{\mathrm{gl},i}^\mathrm{rx}=\arcsin{\left((2i/N_\mathrm{t})+\sin{(\overline{\theta})}\right)}
\end{equation}
for $i \in \mathbb{Z} \setminus \{0\}$, as shown in Fig.~\ref{fig:AF_Rx_ang}. This is caused by the \ac{Rx} inter-element spacing $d_\mathrm{r}=N_\mathrm{t}\lambda/2$, which is larger by a factor of $N_\mathrm{t}$ than the conventional half-wavelength spacing. In Section~\ref{sec:CAF_Rx}, we show that these angle ambiguities in the \ac{Rx} array factor are eliminated in the composite \ac{Tx}–\ac{Rx} array factor, which is equivalent to that of a virtual array with $N_\mathrm{t}N_\mathrm{r}$ elements spaced by $\lambda/2$. Consequently, under the proposed \ac{ELAS} design, the $3\,$dB beamwidth is effectively that of a half-wavelength array with $N_\mathrm{t}N_\mathrm{r}$ elements, i.e., $\Delta\theta_{3\mathrm{dB}}^\mathrm{rx} \approx 0.89\lambda/D_\mathrm{r} \approx 2/(N_\mathrm{t}N_\mathrm{r})$ \cite{Richards}.

\subsubsection{Near-field Rx array factor in the range domain} \label{sec:AF_Rx_range}

Similarly, the \ac{NF} \ac{Rx} array factor in the range domain can be evaluated according to the distance sampling method proposed in \cite{CuiDai22}. This method eliminates the dependence of the \ac{NF} array factor on the \ac{AoA}, by fixing a generic observation direction, i.e., $\theta=\overline{\theta}$. It follows that
\begin{equation} \label{eqn:AF_Rx_range}
    \mathrm{AF}_\mathrm{r}(r) = \frac{1}{N_\mathrm{r}} \sum_{n=-\frac{N_\mathrm{r}-1}{2}}^{\frac{N_\mathrm{r}-1}{2}} e^{-\jmath\pi N_\mathrm{t}^2\frac{\lambda}{4}\cos^2{(\overline{\theta})} n^2 \left(\frac{1}{r}-\frac{1}{\overline{r}}\right)}.
\end{equation}

In contrast to the angular sampling case, the quadratic phase term in \eqref{eqn:AF_Rx_range} makes it difficult to obtain a simple closed-form expression. However, \cite{CuiDai22} shows that $\mathrm{AF}_\mathrm{r}(r)$ can be very accurately approximated in terms of Fresnel functions as
\begin{equation} \label{eqn:AF_Rx_range_approx}
    \mathrm{AF}_\mathrm{r}(r) \approx F(\Xi)=\frac{C(\Xi)+\jmath S(\Xi)}{\Xi}
\end{equation}
where $\Xi=N_\mathrm{r}N_\mathrm{t}\cos{(\overline{\theta})}\sqrt{\frac{\lambda}{8}\left| \frac{1}{r}-\frac{1}{\overline{r}}\right|}$, while $C(\Xi)=\int_0^\Xi\cos{\left(\frac{\pi}{2}t^2\right)dt}$ and $S(\Xi)=\int_0^\Xi\sin{\left(\frac{\pi}{2}t^2\right)dt}$ are the Fresnel functions \cite{Sherman62}.

\begin{figure}[t]
    \centering
    \includegraphics[width=0.83\columnwidth]{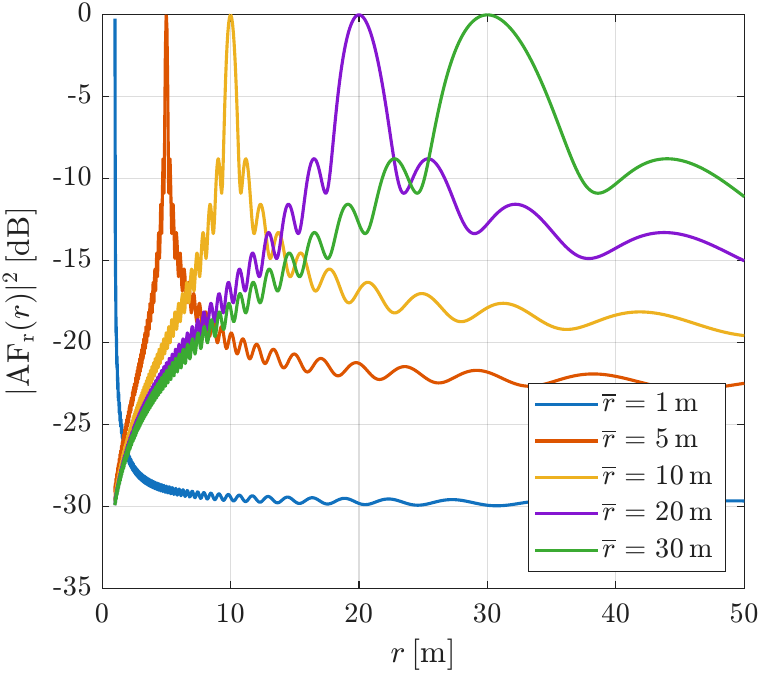}
    \caption{Rx NF array factor in the range domain under the distance sampling method assumption \cite{CuiDai22}, obtained focusing at multiple distances $\overline{r}$ from the transceiver at $\overline{\theta}=0^\circ$ with $f_\mathrm{c}=60\,$GHz, $d_\mathrm{t}=\lambda/2$, $d_\mathrm{r}=N_\mathrm{t}\lambda/2$ and $N_\mathrm{t}=N_\mathrm{r}=32$.}
    \label{fig:AF_Rx_range}
\end{figure}

Note that the profile of $\mathrm{AF}_\mathrm{r}(r)$ depends on the focusing range $\overline{r}$ through the parameter $\Xi$, as illustrated in Fig.~\ref{fig:AF_Rx_range}. In particular, the range resolution determined solely by the \ac{NF} conditions at the \ac{ELAA} \ac{Rx} varies with $\overline{r}$, becoming significantly finer for targets located closer to the array.

\subsection{Composite Array Factor} \label{sec:CAF_Rx}

Referring to the channel model in \eqref{eqn:H}, let us consider a single scatterer, i.e., $P=1$, with \ac{AoD}/\ac{AoA} $\overline{\theta}$ and range $\overline{r}$. The corresponding \ac{MIMO} channel contribution is proportional to the outer product of the transmit and receive array responses, i.e., $\mathbf{H}(\overline{\theta},\overline{r}) \propto \mathbf{b}(\overline{\theta},\overline{r})\,\mathbf{a}^\mathrm{H}(\overline{\theta})$. When transmit beam steering with $\mathbf{w}_\mathrm{t}=\mathbf{a}(\overline{\theta})$ and receive beam focusing with $\mathbf{w}_\mathrm{r}=\mathbf{b}(\overline{\theta},\overline{r})$ are applied, the effective scalar response toward a generic point $(\theta,r)$ is given by
$$  \mathbf{w}_\mathrm{r}^\mathrm{H}\mathbf{H}(\theta,r)\mathbf{w}_\mathrm{t}
    \propto \bigl(\mathbf{a}^\mathrm{H}(\theta)\mathbf{w}_\mathrm{t}\bigr)
            \bigl(\mathbf{w}_\mathrm{r}^\mathrm{H}\mathbf{b}(\theta,r)\bigr).
$$
After normalization, the two factors in parentheses correspond exactly to the transmit and receive array factors defined in \eqref{eqn:AF_Tx} and \eqref{eqn:AF_Rx}. Under the assumptions of the angular sampling method, this motivates defining the normalized composite array factor in the angular domain as
\begin{align} \label{eqn:CAF} \nonumber
    \mathrm{CAF}(\theta) &= \mathrm{AF}_\mathrm{t}(\theta)\mathrm{AF}_\mathrm{r}(\theta)\\
    &= \frac{1}{N_\mathrm{t}N_\mathrm{r}} \frac{\sin{\left(N_\mathrm{r}N_\mathrm{t}\pi/2\left(\sin{(\theta)}-\sin{(\overline{\theta})}\right)\right)}}{\sin{\left(\pi/2\left(\sin{(\theta)}-\sin{(\overline{\theta})}\right)\right)}}.
\end{align}
By comparing \eqref{eqn:CAF} with the \ac{Tx} array factor in \eqref{eqn:AF_Tx}, it is clear that the composite array factor has the same form as the beampattern of a virtual \ac{ULA} with $N_\mathrm{t}N_\mathrm{r}$ elements spaced by $\lambda/2$, thus providing a much larger effective aperture and finer angular resolution, determined by the \ac{Rx} array as $\Delta \theta_{3\mathrm{dB}} \equiv \Delta\theta_{3\mathrm{dB}}^\mathrm{rx} \approx 2/(N_\mathrm{t}N_\mathrm{r})$ (see Section~\ref{sec:AF_Rx}).

As illustrated in Fig.~\ref{fig:CAF_ang}, this behavior is due to the fact that the proposed \ac{Rx} spacing $d_\mathrm{r}=N_\mathrm{t}\lambda/2$ aligns the \ac{Rx} grating lobes with the \ac{Tx} notches. Indeed, from \eqref{eqn:notches_Tx} and \eqref{eqn:gl_Rx} it follows that $\theta_{\mathrm{gl},i}^\mathrm{rx} \equiv \theta_{\mathrm{notch},i}^\mathrm{tx}$ for $i \in \mathbb{Z} \setminus \{0\}$.

In the range domain, the composite array factor coincides with the \ac{Rx} array factor $\mathrm{AF}_\mathrm{r}(r)$. In fact, for the \ac{FF} \ac{Tx} model adopted in this work, the transmit array factor depends only on the angle and not on the range. When we fix the observation angle to the steering direction, i.e., $\theta=\overline{\theta}$, the normalized \ac{Tx} array factor evaluates to $\mathrm{AF}_\mathrm{t}(\overline{\theta})=1$, so that
\begin{equation}
    \mathrm{CAF}(r) = \mathrm{AF}_\mathrm{t}(\overline{\theta})\,\mathrm{AF}_\mathrm{r}(r)
    = \mathrm{AF}_\mathrm{r}(r).
    \label{eqn:composite_range}
\end{equation}
Using the distance sampling method previously introduced, this can be further written as per \eqref{eqn:AF_Rx_range_approx}.

\section{Joint Beam Focusing and Bandwidth Effects for Enhanced Range Resolution} \label{sec:NF_B_r_res}

In the previous section, the composite array response of the proposed \ac{MIMO} architecture was characterized by focusing on its angular and range behavior under a narrowband assumption. 
We now investigate how the range response related to the sole array interacts with the waveform bandwidth in a wideband OFDM system, and how their joint effect determines the achievable range resolution. In particular, in this section, we show that \ac{NF} propagation at the \ac{Rx} can create a super-resolution region in which the array-induced range resolution surpasses the classical bandwidth-limited one, while the bandwidth still plays a key role in controlling sidelobe levels in the overall range profile.
Effective \ac{NF} conditions can significantly enhance the sensing range resolution, even in wideband systems. In an \ac{NF} sensing system, the range resolution is determined by the joint effect of the waveform bandwidth and the \ac{NF} beam focusing operation enabled by \acp{ELAA}. To highlight this interaction, we first recall the classical bandwidth-limited range profile.

For the considered monostatic configuration, the bandwidth $B$ impacts the range resolution according to the Rayleigh criterion as $\Delta r_{\mathrm{B}} = \frac{c}{2B}$, where $\Delta r_{\mathrm{B}}$ is defined as the distance between the main-lobe peak and the first null of the range profile. 
For the considered \ac{OFDM} system, we define the frequency steering vector for a target at range $r$ as
\begin{equation}
    \mathbf{g}(r) =
        [e^{-\jmath 2\pi f_0\frac{2r}{c}}, \dots,e^{-\jmath 2\pi f_{k}\frac{2r}{c}}, \dots,
        e^{-\jmath 2\pi f_{K-1}\frac{2r}{c}}]^\transp.
\end{equation}
When focusing at range $\overline{r}$, the corresponding weighting vector is $\mathbf{w}_\mathrm{B} = \mathbf{g}(\overline{r})$, and the normalized bandwidth-only range profile, by removing the common carrier term from $f_k$, can be expressed as
\begin{align} \label{eqn:RP_B} \nonumber
    \mathrm{RP}_\mathrm{B}(r) & = \frac{1}{K}\,\mathbf{w}_\mathrm{B}^\mathrm{H}\mathbf{g}(r)
    = \frac{1}{K}\,\mathbf{g}^\mathrm{H}(\overline{r})\mathbf{g}(r) \nonumber \\
    &= \frac{1}{K}\sum_{k=-\frac{K-1}{2}}^{\frac{K-1}{2}}
       e^{-\jmath2\pi k\Delta f\frac{2(r-\overline{r})}{c}} \nonumber\\[2pt]
    &= \frac{1}{K}
       \frac{\sin\bigl(K\Delta f\, 2\pi(r-\overline{r})/c\bigr)}
            {\sin\bigl(\Delta f\, 2\pi(r-\overline{r})/c\bigr)}.
\end{align}

On the other hand, the \ac{NF} beam focusing effect also contributes to the range resolution, often described in terms of the \ac{DF} of the array \cite{Liu23}. In this work, we follow the common convention in the \ac{NF} literature and characterize the \ac{DF} by the width of the range interval over which the power of the \ac{NF} array factor remains above a fraction of its peak value. Specifically, under the distance sampling method, we define the \ac{DF} through the condition
\begin{equation} \label{eqn:AF_res}
    \bigl|\mathrm{CAF}(\overline{r}+\Delta r)\bigr|^2 = \bigl|\mathrm{AF}_\mathrm{r}(\overline{r}+\Delta r)\bigr|^2 \;\ge\; \frac{1}{2}.
\end{equation}
Using the Fresnel-function approximation in \eqref{eqn:AF_Rx_range_approx}, the condition in \eqref{eqn:AF_res} can be rewritten as
\begin{equation}
    |F(\Xi)|^2 \;\ge\; \frac{1}{2}
\end{equation}
which is equivalent to $|\Xi| \le \Xi_{3\mathrm{dB}}$ \cite{CuiDai22}, where $\Xi_{3dB}$ is such that $|F(\Xi_{3\mathrm{dB}})|^2 = 0.5$ and is approximately equal to $\Xi_{3\mathrm{dB}} \approx 1.318$. This leads to the constraint $\left|\tfrac{1}{r}-\tfrac{1}{\overline{r}}\right| \le \tfrac{1}{r_\mathrm{DF,r}}$, where, $r_\mathrm{DF,r}$ is commonly referred to as the effective \ac{NF} (or maximum focusing) distance with respect to the \ac{Rx}. For the proposed setup with $d_\mathrm{r} = N_\mathrm{t}\lambda/2$, the latter is given by
\begin{equation}
    r_\mathrm{DF,r}
    \approx \frac{N_\mathrm{r}^2 d_\mathrm{r}^2 \cos^2(\overline{\theta})}{2\lambda\,\Xi_{3\mathrm{dB}}^2}
    = \frac{N_\mathrm{r}^2 N_\mathrm{t}^2 \lambda \cos^2(\overline{\theta})}{8\,\Xi_{3\mathrm{dB}}^2}.
    \label{eqn:rdf_def}
\end{equation}
By recalling \eqref{eq:r_max}, for the proposed \ac{Rx} \ac{ELAA}, the Fraunhofer distance is
$D_\mathrm{F}^\mathrm{rx} = \frac{N_\mathrm{t}^2 (N_\mathrm{r}-1)^2 \lambda}{2}$.
For $\overline{\theta}$ close to broadside (so that $\cos^2(\overline{\theta})\approx 1$) and large $N_\mathrm{r}$,  and using the value of $\Xi_{3\mathrm{dB}}$ given above, \eqref{eqn:rdf_def} can be rewritten as\footnote{The values of $\Xi_{3\mathrm{dB}}$ and, consequently, $r_\mathrm{DF,r}$ are derived according to \cite{WachPollin25} for a \ac{SIMO} \ac{ULA}, i.e., considering only the receive array and ignoring the transmit contribution. This is consistent with the present architecture with dissimilar element spacings, where the \ac{Tx} is assumed to operate in the \ac{FF} and thus does not affect the distance-domain focusing behavior. The reader is referred to \cite{WachPollin25} for a more comprehensive treatment of $\Xi_{3\mathrm{dB}}$ and $r_\mathrm{DF,r}$ across different array architectures.}
\begin{equation} \label{eqn:r_DF}
    r_\mathrm{DF,r}
    \approx \frac{N_\mathrm{r}^2}{4\,\Xi_{3\mathrm{dB}}^2 (N_\mathrm{r}-1)^2}\,D_\mathrm{F}^\mathrm{rx}
    \;\approx\; \frac{D_\mathrm{F}^\mathrm{rx}}{6.952}.
\end{equation}
Solving the inequality above for $r = \overline{r}+\Delta r$ yields \cite{Liu23,WangBjo25}
\begin{equation}
    \Delta r \in
    \left[
        -\frac{\overline{r}^2}{r_\mathrm{DF,r}+\overline{r}},
        \,
        \frac{\overline{r}^2}{\max\bigl(r_\mathrm{DF,r}-\overline{r},\,0\bigr)}
    \right].
\end{equation}
The interval above defines the range span around $\overline{r}$ where the power of the \ac{NF} array factor remains above $0.5$ of its peak, and its width is therefore interpreted as the \ac{DF} of the array.
Thus, beam focusing is effectively achievable only within a limited fraction of the classical Fraunhofer distance, and the impact of the \ac{NF} on the range resolution is confined to this focusing region. In particular, the \ac{DF} of the \ac{Rx} when performing beam focusing in direction $\overline{\theta}$ and at a distance $\overline{r}$ is given by \cite{Liu23}
\begin{equation} \label{eqn:DF}
    \mathrm{DF}_\mathrm{r}(\overline{r})=
    \begin{cases}
        \frac{2\overline{r}^2r_\mathrm{DF,r}}{r_\mathrm{DF,r}^2-\overline{r}^2} & \overline{r} < r_\mathrm{DF,r}\\
        \infty &\overline{r} \ge r_\mathrm{DF,r}.
    \end{cases}
\end{equation}
It is worth noting from \eqref{eqn:DF} that the \ac{DF} depends on the range $\overline{r}$ of the focused location. In particular, it tends to infinity if the focus distance $\overline{r}$ is larger than the threshold $r_\mathrm{DF,r}$, i.e., the focused location is outside the effective \ac{NF} region. This implies that beam focusing degenerates to beam steering, since the orthogonality in the distance domain is almost lost. On the other hand, the range resolution greatly enhances when targets approach the \ac{Rx} \ac{ELAA}, as shown in Fig.~\ref{fig:AF_Rx_range}.

From \eqref{eqn:AF_Rx_range_approx}, \eqref{eqn:composite_range}, and \eqref{eqn:RP_B}, the overall range profile, taking into account both the bandwidth and the beam focusing effects, can be evaluated under the distance sampling method assumptions as
\begin{align} \label{eqn:RP_B_NF} \nonumber
    \mathrm{RP}(r)
    &= \mathrm{RP}_\mathrm{B}(r)\,\mathrm{CAF}(r) \\[2pt]
    &\approx \frac{1}{K}
       \frac{\sin\bigl(K\Delta f\, 2\pi(r-\overline{r})/c\bigr)}
            {\sin\bigl(\Delta f\, 2\pi(r-\overline{r})/c\bigr)}
       \, F\bigl(\Xi(r)\bigr).
\end{align}

\begin{figure} [t]
    \centering
    \includegraphics[width=0.9\columnwidth]{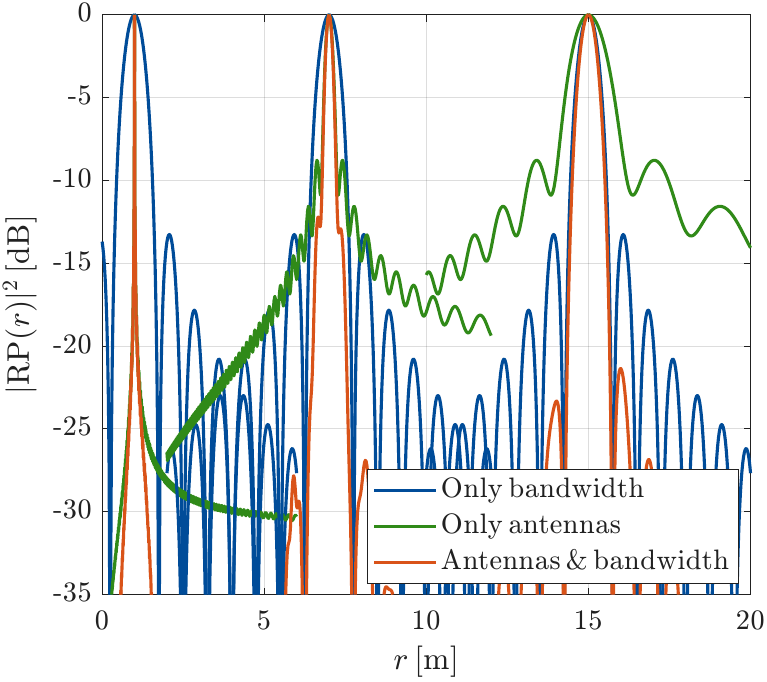}
    \caption{Range profile obtained focusing locations at different distances ($\overline{r}=1,7,15\,$m) from the transceiver at $\overline{\theta}=0^\circ$. Note that, with $f_\mathrm{c}=60\,$GHz, $B=200\,$MHz, $K=1024\,$, $d_\mathrm{t}=\lambda/2$, $d_\mathrm{r}=N_\mathrm{t}\lambda/2$ and $N_\mathrm{t}=N_\mathrm{r}=32$, the effective NF region and super-resolution area are limited by $r_\mathrm{DF,r}=376.1\,$m and $r_\mathrm{sr}=11.9\,$m, respectively.}
    \label{fig:antennas_bandwidth_r_res_mono}
\end{figure}

As shown in Fig.~\ref{fig:antennas_bandwidth_r_res_mono}, when focusing at distances $\overline{r}$ relatively close to the \ac{Rx}, the overall range resolution of the system is determined by $\mathrm{DF}_\mathrm{r}(\overline r)$, which is smaller than the bandwidth-limited Rayleigh range resolution $\Delta r_\mathrm{B}$, and is therefore significantly enhanced compared to conventional radars. In contrast, as the focusing distance $\overline{r}$ increases and $\mathrm{DF}_\mathrm{r}(\overline r)$ becomes larger than $\Delta r_\mathrm{B}$, the overall range resolution is dominated by the bandwidth and coincides with $\Delta r_\mathrm{B}$.
This defines a \emph{super-resolution region} around the \ac{Rx}, where the range resolution is improved beyond the classical bandwidth-limited resolution due to the dominant contribution of the \ac{NF}. This super-resolution region is contained within the effective \ac{NF} area and is bounded by the distance $r_\mathrm{sr}$ from the \ac{Rx}, defined as (in a monostatic setup)
\begin{equation} \label{eqn:super_res}
    r_\mathrm{sr} \triangleq \sqrt{\frac{\Delta r_\mathrm{B}r_\mathrm{DF,r}^2}{2r_\mathrm{DF,r}+\Delta r_\mathrm{B}}} = \sqrt{\frac{r_\mathrm{DF,r}^2}{4Br_\mathrm{DF,r}/c+1}}
\end{equation}
with $r_\mathrm{sr} \le r_\mathrm{DF,r}$. Note that the size of the super-resolution region strongly depends on the system bandwidth. For a fixed number of antennas (and hence fixed $r_\mathrm{DF,r}$), the wider the bandwidth, the smaller the super-resolution region.

Combining the \ac{DF} in \eqref{eqn:DF} with the definition of the super-resolution region in \eqref{eqn:super_res}, the overall range resolution of the monostatic wideband \ac{NF} sensing system as a function of the focusing distance $\overline{r}$ can be written as\footnote{It is worth noting that the overall range resolution expression in \eqref{eqn:NF_res} remains meaningful even though the two contributing terms are defined using different criteria (Rayleigh for $\Delta r_\mathrm{B}$ and half-power width for $\mathrm{DF}_\mathrm{r}$). In fact, as shown in \cite{Cheney09}, when a rectangular transmit pulse $g_\mathrm{tx}(t)$ is used, the Rayleigh and half-power criteria yield approximately equivalent range resolutions.}
\begin{equation} \label{eqn:NF_res}
    \Delta r(\overline{r})=
    \begin{cases}
        \frac{2\overline{r}^2r_\mathrm{DF,r}}{r_\mathrm{DF,r}^2-\overline{r}^2} & \overline{r} < r_\mathrm{sr} \le r_\mathrm{DF,r}\\[4pt]
        \Delta r_\mathrm{B} = \frac{c}{2B} &\overline{r} \ge r_\mathrm{sr}.
    \end{cases}
\end{equation}
In other words, the range resolution is governed by \ac{NF} beam focusing within the super-resolution region ($\overline r < r_\mathrm{sr}$), whereas it is limited by the bandwidth outside this region, even when focusing within the effective \ac{NF} region, i.e., $\overline{r} < r_\mathrm{DF,r}$.

From Fig.~\ref{fig:antennas_bandwidth_r_res_mono}, we also observe that the bandwidth affects the range profile even when the target lies in the super-resolution region. In this regime, the main-lobe width is mainly governed by \ac{NF} beam focusing, whereas the bandwidth helps suppress sidelobes in the overall range profile, especially near the edge of the super-resolution area \cite{Wach2025}. Similarly, for focusing distances within the effective \ac{NF} region but outside the super-resolution region, the range resolution is clearly determined by the bandwidth, while the \ac{NF} array factor still contributes to significant sidelobe attenuation. This shows that, in a wideband \ac{NF} sensing system, bandwidth and \ac{NF} effects not only compete in setting the achievable range resolution, but also jointly improve the overall range profile by reducing the peak sidelobe level.

\section{Target Parameter Detection and Estimation} \label{sec:estimation}

In this work, large-scale arrays are assumed at both \ac{Tx} and \ac{Rx}, so each beam pattern approximately forms a narrow lobe around a steering direction $\overline{\theta}$. Referring to the system model of Section~\ref{sec:system_model}, we adopt a conventional transmit beamforming vector $\mathbf{w}_\mathrm{t} = \mathbf{a}(\overline{\theta})/\|\mathbf{a}(\overline{\theta})\|$. From the \ac{Rx} perspective, the effective channel from the \ac{Tx} to a generic micro-scatterer $p$ of the \ac{ET} is then modulated by the complex beamforming gain $g_p \triangleq \mathbf{a}^\mathrm{H}(\theta_p)\mathbf{w}_\mathrm{t}$, which captures the alignment between the scatterer \ac{AoD} $\theta_p$ and the current transmit beam. For notational convenience, this gain is absorbed into the complex channel coefficient as $h_p \triangleq g_p\,\epsilon_p$, so that, unlike \eqref{eqn:H}, the received signal model is expressed in terms of $h_p$ rather than $\epsilon_p$.

For later processing, it is convenient to rewrite the received signal in \eqref{eq:Rx_signal} in vector form as $\underline{\mathbf{y}} \in \mathbb{C}^{N_\mathrm{r}KM \times 1}$ by stacking the samples across subcarriers, \ac{OFDM} symbols, and antennas
\begin{equation}
    \label{eq:Rx_signal_2}
    \underline{\mathbf{y}}
    = \sum_{p=1}^{P}\sqrt{\frac{P_\mathrm{t}G_\mathrm{t}G_\mathrm{r}}{K}}\,h_p\,
      \mathbf{G}(r_p,\tau_p,\theta_p,f_{\mathrm{D},p})\,\underline{\mathbf{x}}
      + \underline{\boldsymbol{\nu}}
\end{equation}
where $\underline{\mathbf{x}}\in\mathbb{C}^{KM\times1}$ stacks the transmitted symbols $x[k,m]$, $\underline{\boldsymbol{\nu}}\in \mathbb{C}^{N_\mathrm{r}KM \times 1}$ stacks the AWGN samples $\nu_n[k,m]$, and $\mathbf{G}(\cdot)\in \mathbb{C}^{N_\mathrm{r}KM \times KM}$ is the effective channel matrix for scatterer $p$, defined as
\begin{equation}
    \mathbf{G}(r_p,\tau_p,\theta_p,f_{\mathrm{D},p})
    \triangleq
    \mathbf{T}(\tau_p,f_{\mathrm{D},p}) \otimes
    \mathbf{b}(\theta_p,r_p)
\end{equation}
with
\begin{align} \nonumber 
\mathbf{T}(\tau_p,f_{\mathrm{D},p}) \triangleq \mathrm{diag} \Bigl([&1,\dots,e^{\jmath2\pi (M-1)T_\mathrm{s}f_{\mathrm{D},p}}]^\mathrm{T}\otimes\\ 
&\otimes [1,\dots,e^{-\jmath2\pi (K-1)\Delta f \tau_p}]^\mathrm{T} \Bigr) 
\end{align}
a diagonal matrix of size $KM \times KM$.
Since the two-way propagation delay of scatterer $p$ satisfies $\tau_p = 2r_p/c$, we can equivalently write $\mathbf{T}(\tau_p,f_{\mathrm{D},p}) \equiv \mathbf{T}(r_p,f_{\mathrm{D},p})$ and, correspondingly, $\mathbf{G}(r_p,\tau_p,\theta_p,f_{\mathrm{D},p}) \equiv \mathbf{G}(r_p,\theta_p,f_{\mathrm{D},p})$.

For an ET with $P$ micro-scatterers, the unknown parameter set is
$\boldsymbol{\Theta} = \{\boldsymbol{\Theta}_p\}_{p=1}^{P}$, where
$\boldsymbol{\Theta}_p = \{|h_p|,\angle h_p,r_p,\theta_p,f_{\mathrm{D},p}\}$.
The joint \ac{ML} estimate involves a search over
$\Gamma \triangleq \mathbb{C}^P \times \mathbb{R}^{3P}$ \cite{dehkordi2023multistatic}
\begin{align} \label{eqn:ML}
    \boldsymbol{\Theta}_{\mathrm{ML}}
    = \underset{\boldsymbol{\Theta} \in \Gamma}{\arg\min}\;
      \left\| \underline{\mathbf{y}}
      - \sum_{p=1}^{P}\sqrt{\frac{P_\mathrm{t}G_\mathrm{t}G_\mathrm{r}}{K}}\,
      h_p\, \mathbf{G}(r_p,\theta_p,f_{\mathrm{D},p}) \,\underline{\mathbf{x}}
      \right\|^2.
\end{align}
However, solving \eqref{eqn:ML} would require knowledge of the number of scatterers $P$ and leads to a high-dimensional model-order selection problem \cite{MarGioChi:J15sp}, which is intrinsically ill-posed for ETs with many indistinguishable micro-scatterers \cite{Rife}.
Following the grid-based \ac{GLRT} framework adopted in \cite{dehkordi2023multistatic}, we instead consider a discrete search grid
$\Psi$ over $(r,\theta,f_\mathrm{D})$ and assume that at most one micro-scatterer can lie in each grid cell. 

Starting from the likelihood associated with \eqref{eqn:ML} and substituting the ML estimate of the complex gain $h_p$ (see \cite[eq. (41)]{dehkordi2023multistatic}), the resulting \ac{GLRT} metric for a candidate triplet $(r,\theta,f_\mathrm{D}) \in \Psi$ can be written as
\begin{equation} \label{eqn:GLRT}
\ell(r,\theta,f_\mathrm{D})
= \frac{\bigl|\boldsymbol{\xi}(r,f_\mathrm{D})\,\mathbf{b}^\ast(\theta,r)\bigr|^2}
       {\underline{\mathbf{x}}^\mathrm{H}\,\mathbf{b}^\mathrm{H}(\theta,r)\mathbf{b}(\theta,r)\,\underline{\mathbf{x}}}\underset{\mathcal{H}_0}{\overset{\mathcal{H}_1}{\gtrless}} \eta
\end{equation}
with $\boldsymbol{\xi}(r,f_\mathrm{D}) = \underline{\mathbf{x}}^\mathrm{H}
\mathbf{T}^\mathrm{H}(r,f_{\mathrm{D}})\mathbf{Y}^\mathrm{T}$ and
$\mathbf{Y} = [\mathbf{y}[0,0],\dots,\mathbf{y}[M-1,K-1]] \in \mathbb{C}^{N_\mathrm{r} \times KM}$ obtained by stacking the received vectors in \eqref{eq:Rx_signal} over all subcarriers and OFDM symbols. 

In \eqref{eqn:GLRT}, a binary hypothesis test is performed to decide whether the cell contains a micro-scatterer ($\mathcal{H}_1$) or not ($\mathcal{H}_0$) with the threshold $\eta$ chosen to guarantee a given \ac{FAR}. Under $\mathcal{H}_0$, when only noise is present at the receiver (i.e., $\underline{\mathbf{y}}=\underline{\boldsymbol{\nu}}$), the statistic $\ell(r,\theta,f_\mathrm{D})$ is exponentially distributed with mean $\sigma_\nu^2$. Hence, for a desired per-bin false-alarm probability $P_{\mathrm{FA},\mathrm{point}}$, the threshold is
\begin{equation}
    \eta = -\sigma_\nu^2 \ln\bigl(P_{\mathrm{FA},\mathrm{point}}\bigr).
    \label{eq:thr}
\end{equation}
To guarantee a given \ac{FAR} over the entire search grid $\Psi$ with
$\mathrm{card}(\Psi)$ points, we set $P_{\mathrm{FA},\mathrm{point}} = \mathrm{FAR} / \mathrm{card}(\Psi)$.

The outcome of the procedure in \eqref{eqn:GLRT} is a \emph{radar image} in the $(r,\theta,f_\mathrm{D})$ space, where localized peaks of $\ell(r,\theta,f_\mathrm{D})$ correspond to dominant scattering points of the \ac{ET}, in line with the interpretation in \cite{dehkordi2023multistatic}. In practice, this radar map is obtained through a coordinated beam-scanning procedure. The \ac{Tx} beam is steered over a discrete set of $N_\mathrm{d}$ directions
$\{\overline{\theta}_q\}_{q=0}^{N_\mathrm{d}-1}$ that cover a given \ac{RoI}. For each direction $\overline{\theta}_q$, the transmit beamforming vector is set to $\mathbf{w}_\mathrm{t} = \mathbf{a}(\overline{\theta}_q)$, and the \ac{Rx} performs beam focusing within the corresponding beamspace sector according to the \ac{NF} steering vector $\mathbf{b}(\theta,r)$ used in \eqref{eqn:GLRT}.
The scan step is chosen on the order of the $3\,$dB beamwidth of the \ac{Tx} array, $\Delta\theta_{3\mathrm{dB}}^\mathrm{tx} \approx 2/N_\mathrm{t}$ (in radians), so that adjacent beams overlap and jointly cover the \ac{RoI}. Since the \ac{Tx} and \ac{Rx} arrays are co-located, this transmit–receive scanning can be naturally synchronized, and the \ac{GLRT} metric in \eqref{eqn:GLRT} is evaluated for all $(r,\theta,f_\mathrm{D})$ grid points illuminated by each beam. After all $N_\mathrm{d}$ beams have been processed, the collection of \ac{GLRT} values forms a radar map over the \ac{RoI}. For a more comprehensive explanation, readers can refer to \cite{BacBis25}.

Finally, once the range–angle pair $(\hat{r}_p,\hat{\theta}_p)$ of each detected micro-scatterer is estimated from the grid, its Cartesian position is obtained as
\begin{equation} \label{eqn:pos}
    \hat{\mathbf{p}}_{p}
    = [\hat{x}_p,\hat{y}_p]^\mathrm{T}
    = [\hat{r}_p\cos(\hat{\theta}_p),\,\hat{r}_p\sin(\hat{\theta}_p)]^\mathrm{T}.
\end{equation}

\section{Numerical Results} \label{sec:numerical_results}

\begin{table}[t]
\caption{}
\begin{center}
\resizebox{1\columnwidth}{!}{
\renewcommand{\arraystretch}{0.85}
\vspace{-3mm}
\begin{tabular}{clc}
    \toprule
    \multicolumn{3}{c}{\textbf{Simulation parameters}}\\
    \midrule
    $f_\mathrm{c}$ & Carrier frequency & $60\,$GHz\\
    $K$ & Active subcarriers for sensing & $1024$\\
    $M$ & OFDM symbols & $10$\\
    $N_\mathrm{t}$ / $N_\mathrm{r}$ & Number of Tx / Rx antennas & $32$\\
    $D_\mathrm{t}$ & Tx aperture array & $0.08\,$m\\
    $D_\mathrm{r}$ & Rx aperture array & $2.48\,$m\\
    $G_\mathrm{t}$ / $G_\mathrm{r}$ & Tx / Rx antenna element gain & $1$\\
    $\mathrm{EIRP} \triangleq P_\mathrm{t}G_\mathrm{t}N_\mathrm{t}$ & Equivalent isotropic radiated power & $0.1\,$W\\
    $n_\mathrm{F}$ & Receiver noise figure & $10\,$dB\\
    $\sigma_p=\sigma$ & Micro-scatterer RCS & $0.25$ m$^2$\\
    $A_\mathrm{RoI}$ & RoI area & $3\times3\,$m$^2$\\
    $D_\mathrm{F}^\mathrm{tx}$ & Tx Fraunhofer distance & $2.4\,$m\\
    $D_\mathrm{F}^\mathrm{rx}$ & Rx Fraunhofer distance & $2460\,$m\\
    $L$ / $W$ & ET length / width & $2\,$m / $1\,$m\\
    $q$ & ET grid scatterer probability & $0.1$\\
    $|\mathcal{E}|$ & Number of ET grid elements & $200$\\
    $\mathrm{FAR}$ & Expected number of false-alarm bins per map & $1$\\
    $N_\mathrm{MC}$ & Monte Carlo iterations & $1000$\\
    \bottomrule
\end{tabular}
}
\end{center}
\label{tab:sim_param}
\end{table}
\begin{figure*}[t]
    \centering
    \begin{subfigure}[b]{0.315\textwidth}
        \centering
        \includegraphics[width=\textwidth]{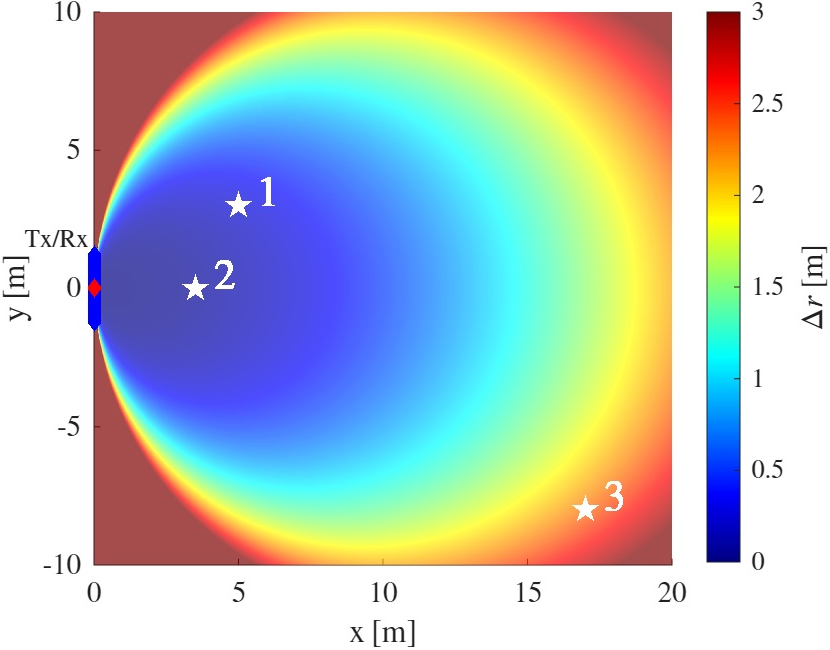}
        \caption{$B=50\,$MHz}
        \label{fig:B_50}
    \end{subfigure}
    \hspace{0.25cm}
    \begin{subfigure}[b]{0.315\textwidth}
        \centering
        \includegraphics[width=\textwidth]{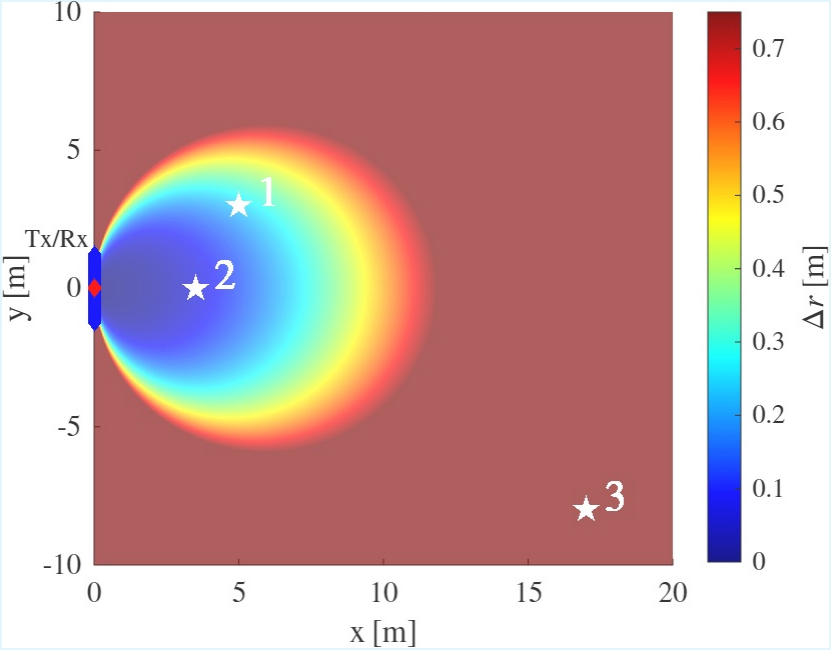}
        \caption{$B=200\,$MHz}
        \label{fig:B_100}
    \end{subfigure}
    \hspace{0.25cm}
    \begin{subfigure}[b]{0.315\textwidth}
        \centering
        \includegraphics[width=\textwidth]{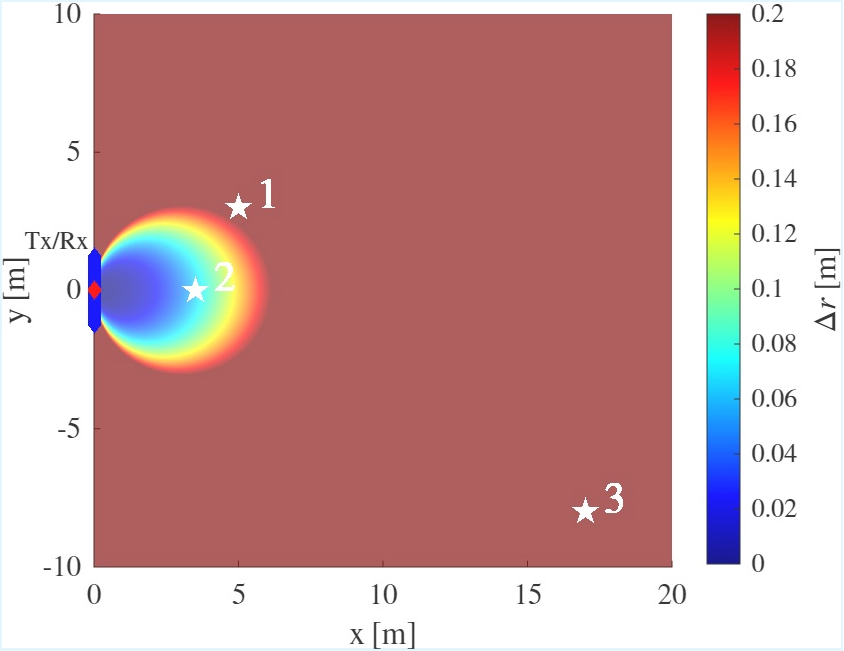}
        \caption{$B=750\,$MHz}
        \label{fig:B_750}
    \end{subfigure}
    \caption{Range-resolution map $\Delta r(x,y)$ over the monitored area for (a) $B=50\,$MHz, (b) $B=200\,$MHz, and (c) $B=750\,$MHz by using the \ac{ELAS} design with $f_\mathrm{c}=60\,$GHz and $N_\mathrm{t}=N_\mathrm{r}=32$. These maps illustrate how the super-resolution region (low $\Delta r$ area) shrinks as the bandwidth increases. White stars labelled 1–3 indicate the ET centroid positions for the three simulation setups: (1) $\mathbf{p}_0=[5,3]^\mathsf{T}$, (2) $\mathbf{p}_0=[3.5,0]^\mathsf{T}$, and (3) $\mathbf{p}_0=[17,-8]^\mathsf{T}$.}
    \label{fig:deltaR_vs_B}
\end{figure*}
Numerical simulations are performed to evaluate the effectiveness of the proposed \ac{ELAS} method and to quantify how the enhanced \ac{NF} range resolution impacts the position-estimation accuracy of the scatterers of an \ac{ET} in a wideband \ac{OFDM}-based monostatic sensing system. The target moves inside a monitored area of size $20 \times 20\,$m$^2$, as shown in Fig.~\ref{fig:deltaR_vs_B}, where the white stars labeled $1$, $2$, and $3$ mark the centroids of the rectangular \ac{ET} in the three simulation setups. In particular, the \ac{ET} centroid is placed at $\mathbf{p}_0=[5,3]^\mathrm{T}$ and $\mathbf{p}_0=[3.5,0]^\mathrm{T}$ in configurations~$1$ and~$2$, with tangential and normal headings with respect to the array, respectively, and at $\mathbf{p}_0=[17,-8]^\mathrm{T}$ in configuration~$3$, where it is also normally oriented.

First, we show that the angle ambiguities caused by the large inter-element spacing at the \ac{Rx} can be naturally avoided when the transmit–receive spacing prescribed by \ac{ELAS} is adopted. Then, we assess localization performance by computing the \ac{GOSPA} metric over the \ac{ET} scatterers and the \ac{RMSE} of the estimated \ac{ET} centroid.

The simulations consider the system parameters in Table~\ref{tab:sim_param} and the \ac{QPSK} constellation for each subcarrier, while the system bandwidth $B$ is varied in the range $50$–$750\,$MHz to investigate its effect on localization resolution and accuracy in wideband \ac{NF} sensing. As $B$ changes, the subcarrier spacing $\Delta f$ (and hence the \ac{OFDM} symbol duration $T_\mathrm{s}$) is adjusted accordingly to keep the number of active subcarriers fixed as per Table~\ref{tab:sim_param}. Notably, with the proposed \ac{ELAS} design and $N_\mathrm{t}=N_\mathrm{r}=32$ antennas at both \ac{Tx} and \ac{Rx}, we obtain a large Fraunhofer distance $D_\mathrm{F}^\mathrm{rx}=2460\,$m and an effective maximum focusing distance $r_\mathrm{DF,r}\approx 376\,$m at the \ac{Rx}, thus achieving an extended \ac{NF} region with a relatively small number of elements. The impact of this maximum focusing distance on the overall system range resolution for different bandwidths is analyzed in detail in the next subsection.

\subsection{Range Resolution of Wideband Near-Field Sensing}

Fig.~\ref{fig:deltaR_vs_B} shows the resulting range-resolution map $\Delta r(x,y)$ over the monitored area for three bandwidth values, $B = \{50, 200, 750\}\,$MHz, obtained from \eqref{eqn:NF_res}. As mentioned above, this area is entirely contained within the effective \ac{NF} region of the \ac{Rx}, since $r_\mathrm{DF,r}\approx 376\,$m for the parameters in Table~\ref{tab:sim_param}. The plots clearly highlight two regimes: (i) a central super-resolution region, where $\Delta r(x,y)$ is dominated by the \ac{DF} term $\mathrm{DF}_\mathrm{r}(\overline r)$ and can be significantly smaller than the bandwidth-limited resolution $\Delta r_\mathrm{B}$, and (ii) an outer region where the range resolution is essentially flat and equal to $\Delta r_\mathrm{B}$.

As the bandwidth decreases, the super-resolution region expands towards
the edge of the monitored area, so that for $B=50\,$MHz it covers almost
the entire $20\times20\,$m$^2$ domain. In this regime, when the system bandwidth is limited, \ac{NF} beam focusing becomes the dominant driver of range resolution. Conversely, for larger bandwidths (e.g., $B=750\,$MHz), the super-resolution region shrinks to a small neighborhood around the array. In this case, the overall range resolution is predominantly set by $\Delta r_\mathrm{B}$ over most of the monitored area, although the entire area still lies within the effective \ac{NF} region defined by $r_\mathrm{DF,r}$. This confirms that, in wideband \ac{NF} sensing, the \ac{DF} alone is not sufficient to improve range resolution beyond the bandwidth limit; \ac{NF} gains are realized only within the super-resolution region, whose spatial extent is governed by both $B$ and $r_\mathrm{DF,r}$ according to \eqref{eqn:super_res}.

\begin{figure}[t]
    \centering
    \includegraphics[width=0.8\columnwidth]{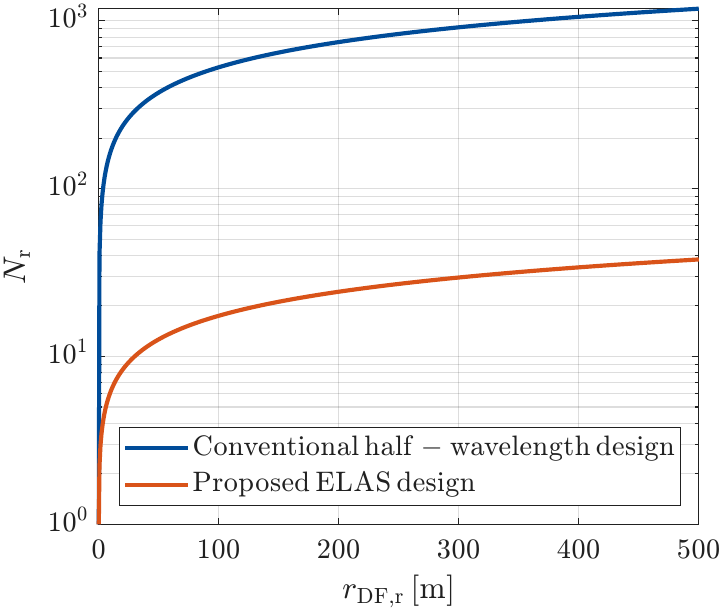}
    \caption{Comparison of the number of \ac{Rx} antenna elements required to achieve a target maximum focusing distance $r_\mathrm{DF,r}$ for the proposed \ac{ELAS} architecture and a conventional half-wavelength-spaced array, with $f_\mathrm{c}=60\,\mathrm{GHz}$ and $N_\mathrm{t}=32$.}
    \label{fig:r_DF_vs_Nr}
\end{figure}

For a fixed bandwidth, the size of the super-resolution region grows with $r_\mathrm{DF,r}$ and thus with the aperture of the receive \ac{ELAA}. With the proposed \ac{ELAS} design, we choose $d_\mathrm{r}=N_\mathrm{t}\lambda/2$ and $d_\mathrm{t}=\lambda/2$, so that the effective aperture of the \ac{Rx} is on the order of $N_\mathrm{r}N_\mathrm{t}\lambda/2$, while only $N_\mathrm{r}$ physical elements are used. In contrast, a conventional $\lambda/2$-spaced \ac{ELAA} would need approximately $N_\mathrm{t}N_\mathrm{r}$ elements to achieve a comparable aperture.
Therefore, \ac{ELAS} achieves a large $r_\mathrm{DF,r}$, resulting in a wide super-resolution region, while reducing the number of antennas and \ac{RF} chains by approximately a factor of $N_\mathrm{t}$, as can be seen in Fig.~\ref{fig:r_DF_vs_Nr}. This approach significantly lowers system complexity and power consumption, yet preserves the \ac{NF} range-resolution gain.

\subsection{Impact of the Enhanced Wideband Near-Field Range Resolution on ET Position Estimation Accuracy}

\begin{figure*}[t]
    \centering
    \begin{subfigure}[b]{0.3\textwidth}
        \centering
        \includegraphics[width=\textwidth]{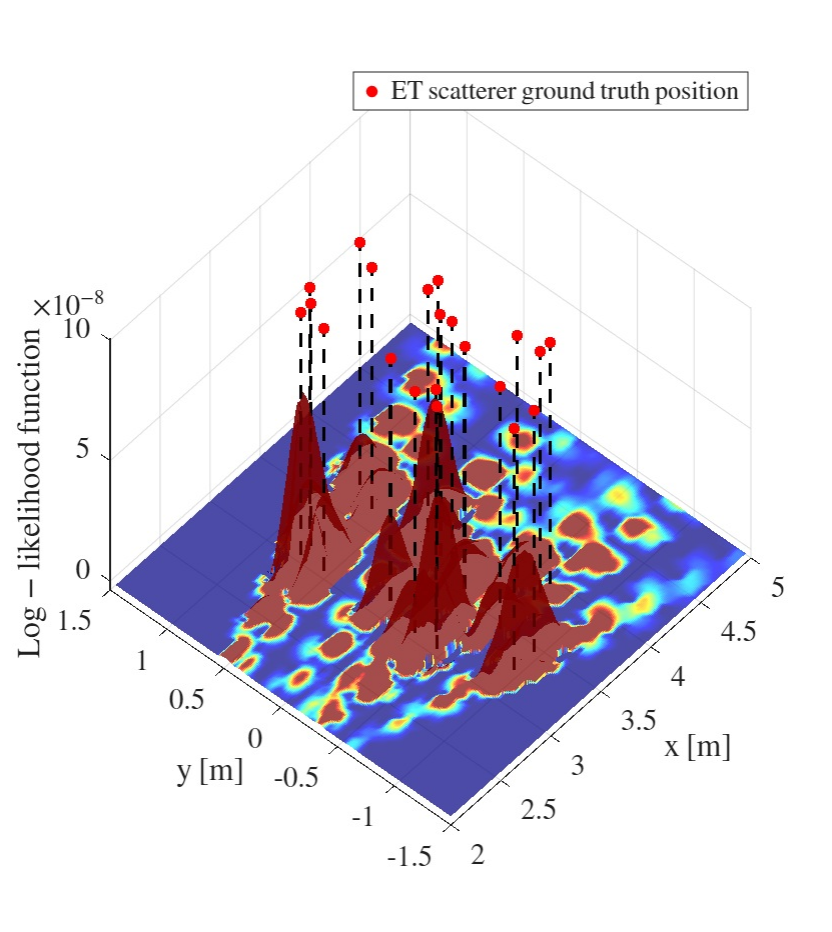}
        \caption{FF (without ELAS) in setup~$2$ ($f_\mathrm{c}=60\,$GHz, $N_\mathrm{t}=N_\mathrm{r}=32$, $d_\mathrm{t}=d_\mathrm{r}=\lambda/2$, $B~=~400\,$MHz)}
        \label{fig:likelihood_FF}
    \end{subfigure}
    \hspace{0.5cm}
    \begin{subfigure}[b]{0.3\textwidth}
        \centering
        \includegraphics[width=\textwidth]{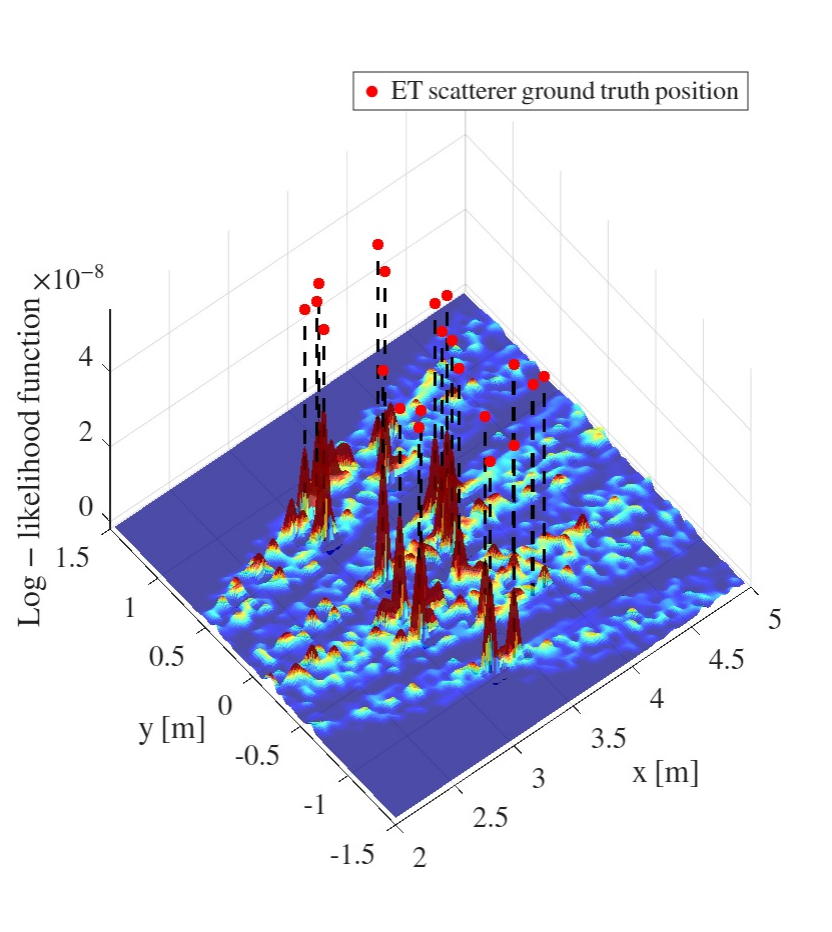}
        \caption{Super-resolution NF (with ELAS) in setup~$2$ ($f_\mathrm{c}=60\,$GHz, $N_\mathrm{t}=N_\mathrm{r}=32$, $d_\mathrm{t}~=~\lambda/2,d_\mathrm{r}=N_\mathrm{t}\lambda/2$, $B~=~50\,$MHz)}
        \label{fig:likelihood_NF_50MHz}
    \end{subfigure}
    \hspace{0.5cm}
    \begin{subfigure}[b]{0.3\textwidth}
        \centering
        \includegraphics[width=\textwidth]{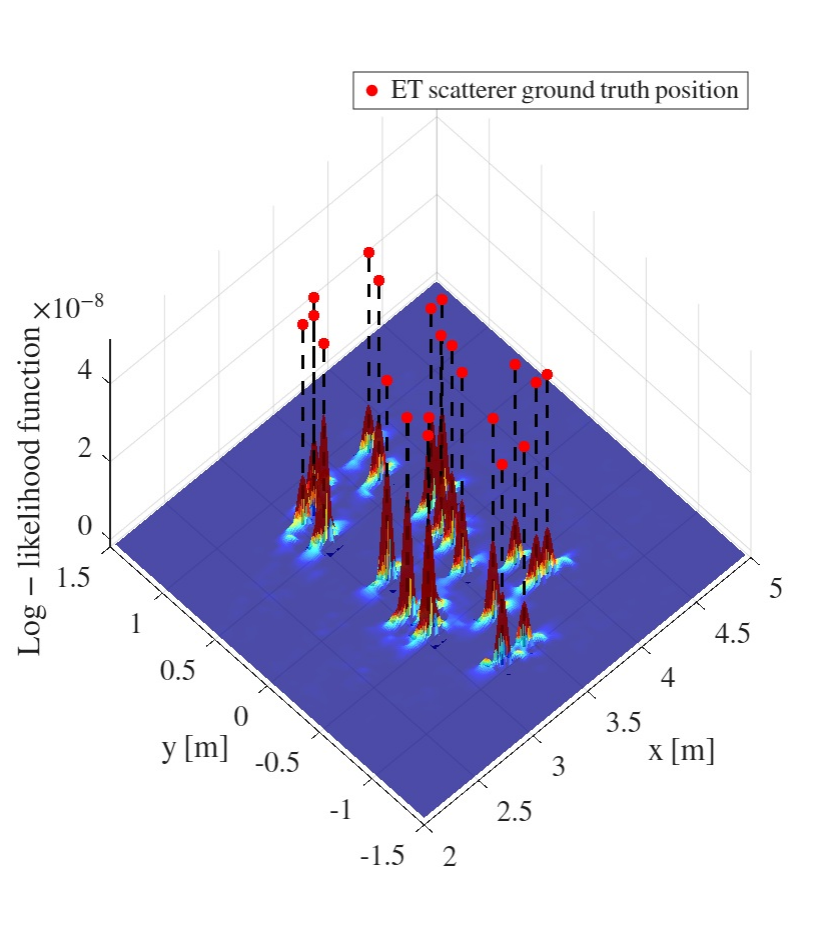}
        \caption{Super-resolution NF (with ELAS) in setup~$2$ ($f_\mathrm{c}=60\,$GHz, $N_\mathrm{t}=N_\mathrm{r}=32$, $d_\mathrm{t}~=~\lambda/2,d_\mathrm{r}=N_\mathrm{t}\lambda/2$, $B~=~400\,$MHz)}
        \label{fig:likelihood_NF_400MHz}
    \end{subfigure}
    \begin{subfigure}[b]{0.3\textwidth}
        \centering
        \includegraphics[width=\textwidth]{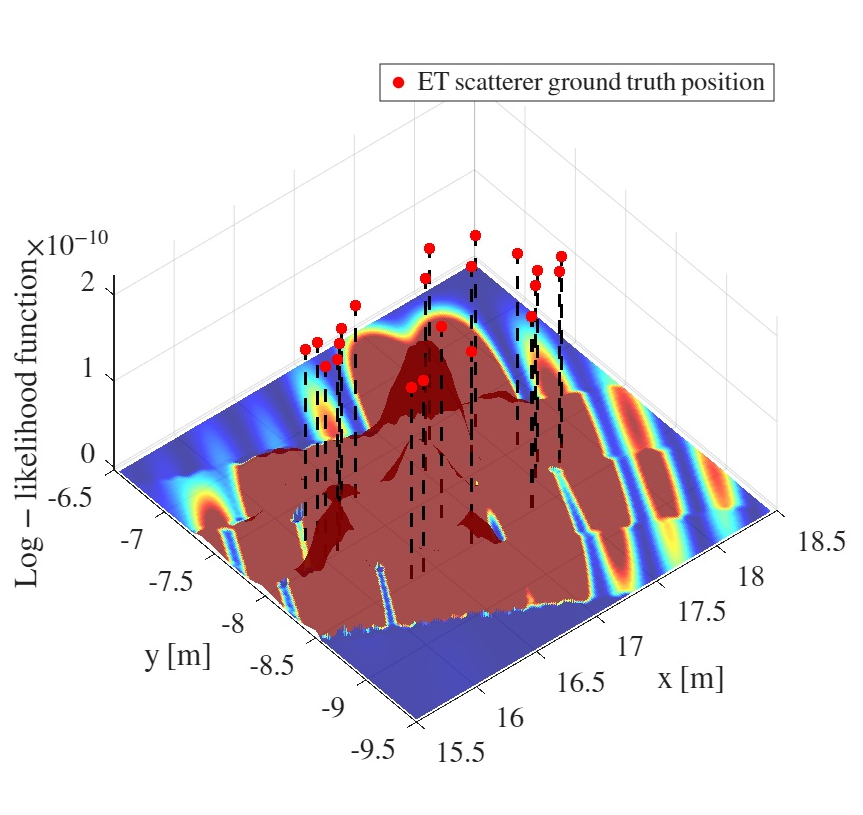}
        \caption{FF (without ELAS) in setup~$3$ ($f_\mathrm{c}=60\,$GHz, $N_\mathrm{t}=N_\mathrm{r}=32$, $d_\mathrm{t}=d_\mathrm{r}=\lambda/2$, $B~=~400\,$MHz)}
        \label{fig:FF_setup3}
    \end{subfigure}
    \hspace{0.5cm}
    \begin{subfigure}[b]{0.3\textwidth}
        \centering
        \includegraphics[width=\textwidth]{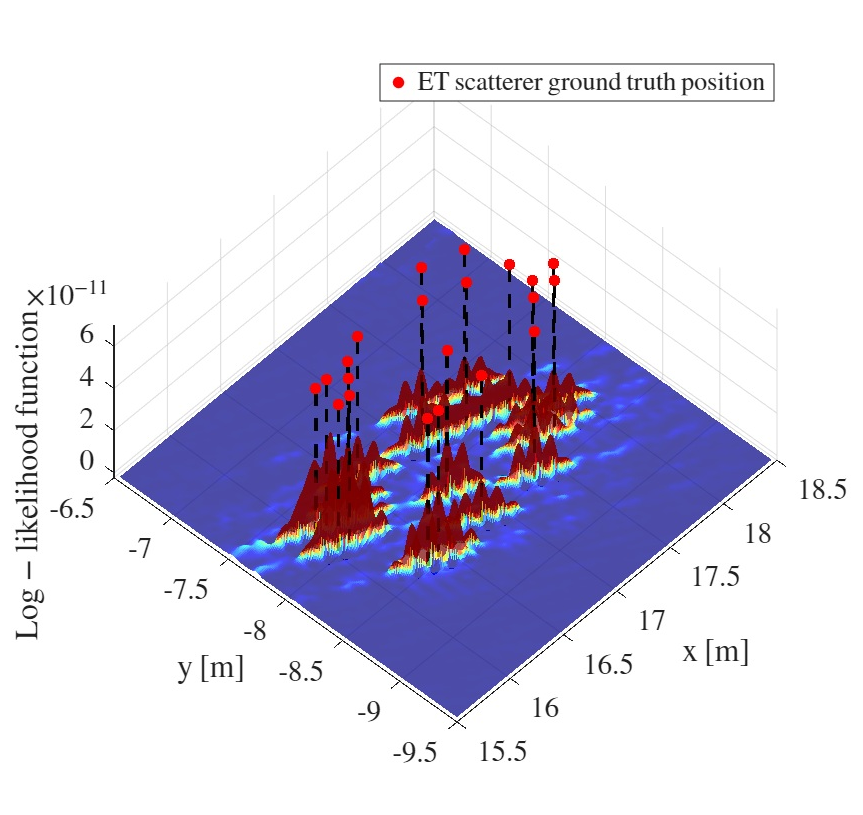}
        \caption{Effective NF (with ELAS) in setup~$3$ ($f_\mathrm{c}=60\,$GHz, $N_\mathrm{t}=N_\mathrm{r}=32$, $d_\mathrm{t}~=~\lambda/2,d_\mathrm{r}=N_\mathrm{t}\lambda/2$, $B~=~400\,$MHz)}
        \label{fig:NF_setup3}
    \end{subfigure}
    \caption{Log-likelihood profile within the search RoI for different simulation setups and system bandwidths, evaluated under both FF conditions with the conventional half-wavelength spacing at both Tx and Rx, and effective NF propagation enabled by the proposed ELAS method.}
    \label{fig:likelihood_profile}
\end{figure*}
In general, range resolution and localization accuracy are distinct performance metrics: the former quantifies the ability to separate closely spaced scatterers, whereas the latter measures how close the estimated positions are to their true locations. For \acp{ET}, however, these metrics are strongly coupled: as the radar range resolution improves, more scattering contributions from the \ac{ET} can be individually resolved, providing richer spatial information. In this sense, higher resolution not only refines the structural representation of the target but also tends to improve the accuracy of its geometric center. 

To quantify this effect, we simulate a wideband \ac{NF} \ac{OFDM}-based \ac{ISAC} system with an \ac{ET} located in configurations $1$, $2$, and $3$ of Fig.~\ref{fig:deltaR_vs_B}, using $N_\mathrm{MC}=1000$ Monte Carlo trials. As discussed in Section~\ref{sec:estimation}, target parameter estimation is performed over a \ac{RoI} of size $3\times3\,$m$^2$ centered at the true centroid position $\mathbf{p}_0$ of the \ac{ET}. The \ac{GLRT} detector in \eqref{eqn:GLRT} is employed for joint detection and estimation, with the threshold set according to \eqref{eq:thr} to guarantee a total \ac{FAR} per radar map equal to $\mathrm{FAR}=1$, i.e., on average one noise-only pixel exceeding the threshold per map.

\begin{figure*}[t]
    \centering
    \begin{subfigure}{0.44\linewidth}
        \centering
        \includegraphics[width=\linewidth]{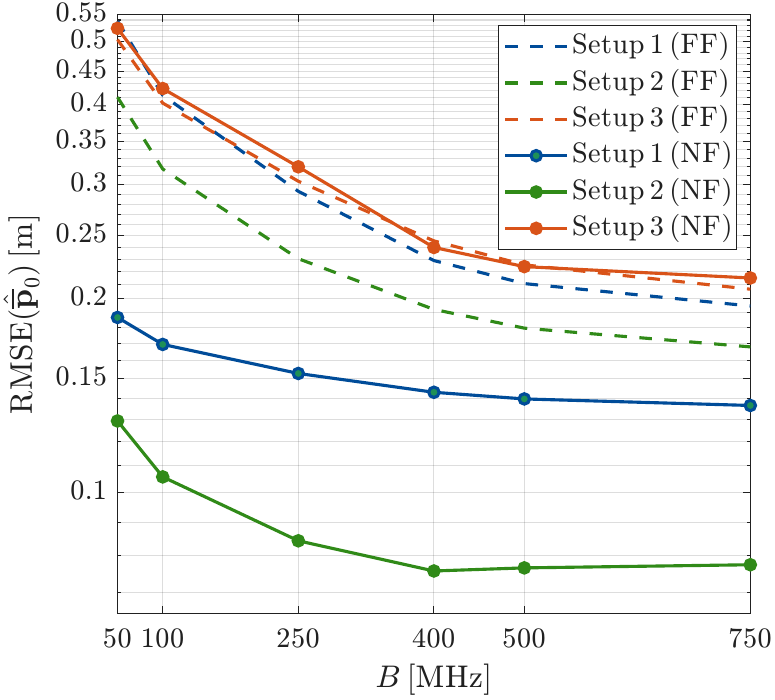}
        \caption{RMSE}
        \label{fig:RMSE_vs_B}
    \end{subfigure}
    \hfill
    \begin{subfigure}{0.44\linewidth}
        \centering
        \includegraphics[width=\linewidth]{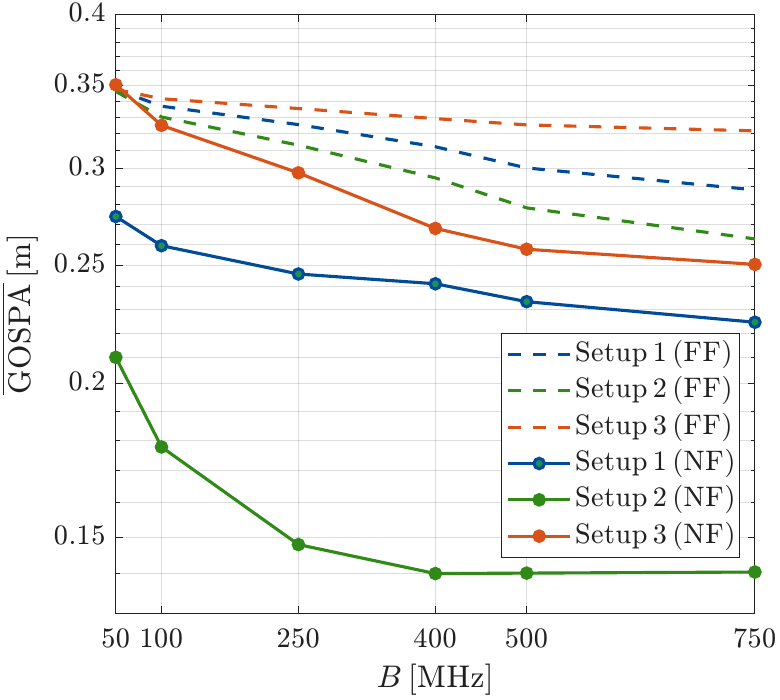}
        \caption{Average GOSPA}
        \label{fig:Average_GOSPA_vs_B}
    \end{subfigure}
    \caption{Localization performance as a function of the system bandwidth $B$ with $N_\mathrm{t}=N_\mathrm{r}=32$ antennas, in FF conditions with conventional half-wavelength spacing ($d_\mathrm{t}=d_\mathrm{r}=\lambda/2$) and in the effective NF of the Rx enabled by the proposed ELAS method ($d_\mathrm{t}=\lambda/2$, $d_\mathrm{r}=N_\mathrm{t}\lambda/2$). (a) RMSE of ET scattering centroid position $\hat{\overline{\mathbf{p}}}_0$; (b) average GOSPA over the detected scatterers.}
    \label{fig:perf_vs_B}
\end{figure*}

In this work, we focus on a stationary \ac{ET}, so its Doppler shift is identically zero over the measurement interval. Accordingly, in the \ac{GLRT} of \eqref{eqn:GLRT}, we fix the Doppler to its true value, $f_\mathrm{D}=0\,$Hz, and evaluate the metric over a two-dimensional grid in $(r,\theta)$: $\ell(r,\theta) \triangleq \ell(r,\theta,f_\mathrm{D}{=}0)$.
This reduces the problem dimensionality by treating the Doppler as a known part of the model (i.e., static target hypothesis). Moreover, in the considered setup, only $M=10$ \ac{OFDM} symbols are available, so the Doppler resolution $1/(M T_\mathrm{s})$ is relatively coarse and would not allow reliable discrimination between a slowly moving \ac{ET} and static objects based on Doppler information alone. In fact, as shown in \cite{BacBis25} for a similar setting, using $f_\mathrm{D}=0\,$Hz as a reference Doppler value yields a negligible impact on localization performance even for slowly moving targets, while significantly reducing the computational burden compared to a full $3$-dimensional search.

The radar map in the Cartesian $x$–$y$ domain is obtained by evaluating $\ell(r,\theta)$ on a polar grid within the \ac{RoI}, applying the \ac{GLRT} detector in \eqref{eqn:GLRT}, and mapping each grid point $(r,\theta)$ to $(x,y)$ via the polar-to-Cartesian transformation in \eqref{eqn:pos}. As illustrated in Fig.~\ref{fig:likelihood_profile}, the resulting maps exhibit multiple local peaks in the $x$–$y$ plane. While some peaks may arise from sidelobes, most are expected to correspond to effective scattering points of the \ac{ET}. The peak locations thus provide estimates of the micro-scatterer positions. To refine detection, a peak-picking procedure is applied to the map so that only local maxima of $\ell(r,\theta)$ are retained as detected points, rather than using the full \ac{GLRT} profile.

The performed simulations analyze how the enhanced range resolution produced by the joint effect of bandwidth and \ac{NF} propagation impacts \ac{ET} localization, both inside and outside the super-resolution region. To this end, we evaluate two localization metrics as functions of the system bandwidth $B$.

The first metric is the \ac{RMSE} of the \ac{ET} scattering centroid estimation, whose ground-truth value for the $i$-th iteration is defined as the barycenter of the scatterer positions
$\overline{\mathbf{p}}_0^{\,i} \triangleq (1/P^{(i)})\sum_{p=1}^{P^{(i)}}\mathbf{p}_p$, where $P^{(i)}$ is the number of \ac{ET} scatterers at a given Monte Carlo run $i$. The corresponding \ac{RMSE} is
    $$\mathrm{RMSE}(\hat{\overline{\mathbf{p}}}_0)
    = \sqrt{\tfrac{1}{N_\mathrm{MC}}\sum_{i=1}^{N_\mathrm{MC}}
      \bigl\|\hat{\overline{\mathbf{p}}}_0^{\,i}-\overline{\mathbf{p}}_0^{\,i}\bigr\|^2}$$
where $\overline{\mathbf{p}}_0^{\,i}$ and $\hat{\overline{\mathbf{p}}}_0^{\,i}$ are the true and estimated centroids at the $i$-th trial, respectively. This metric reflects both the ability to resolve multiple scattering points and the accuracy of their position estimates.

As a second metric, we use the \ac{GOSPA} distance \cite{Rah17}, widely adopted in multi-target and multi-scatterer localization. \ac{GOSPA} compactly captures localization error, missed detections, and false alarms in a single scalar score. We adopt the standard \ac{GOSPA} formulation as per \cite[Eq. (12)]{Mat25}, with gating parameter $\xi_g = 0.5\,\text{m}$ (chosen to be on the order of five times the grid spacing, i.e., the distance between two adjacent pixels in the \ac{RoI}), and order $q=2$, and compute the average \ac{GOSPA} over $N_\mathrm{MC}$ trials. In particular, estimated scatterers farther than $\xi_g$ from any true scatterer are treated as false alarms, and true scatterers without an estimate within $\xi_g$ are counted as missed detections. A lower \ac{GOSPA} value indicates better overall \ac{ET} localization performance.

The impact of \ac{NF} operation is assessed by comparing it with a \ac{FF} baseline under the simulation parameters in Table~\ref{tab:sim_param}. As a \ac{FF} baseline, we consider conventional half-wavelength \acp{ULA} at both the \ac{Tx} and \ac{Rx}, while as an \ac{NF} alternative, we employ the proposed \ac{ELAS} design, so that the \ac{ET} lies in the effective \ac{NF} region of the \ac{Rx}. 

Fig.~\ref{fig:likelihood_profile} illustrates the \ac{GLRT}-based radar maps for representative cases. In Fig.~\ref{fig:likelihood_NF_50MHz} and Fig.~\ref{fig:likelihood_NF_400MHz}, corresponding to setup~$2$ of Fig.~\ref{fig:deltaR_vs_B}, i.e., with the \ac{ET} in the \ac{NF} super-resolution region, the proposed \ac{ELAS} spacing yields ambiguity-free angle profiles despite the large inter-element spacing at the \ac{Rx}. Comparing the \ac{FF} profile in Fig.~\ref{fig:likelihood_FF} (half-wavelength arrays at both ends) with the \ac{NF} profile in Fig.~\ref{fig:likelihood_NF_400MHz} for the same bandwidth $B=400\,$MHz, we clearly observe the enhanced range resolution enabled by \ac{NF} operation, with the scattering contributions from the \ac{ET} much more finely separated than the \ac{FF} case. Moreover, as discussed in Section~\ref{sec:NF_B_r_res}, bandwidth still plays a key role even within the super-resolution region. The comparison between Fig.~\ref{fig:likelihood_NF_50MHz} and Fig.~\ref{fig:likelihood_NF_400MHz} shows that increasing $B$ substantially suppresses sidelobes in the range profile, yielding a cleaner overall radar image. 
The interplay between bandwidth and \ac{NF} beam focusing is also observed outside the super-resolution region, as outlined in Section~\ref{sec:NF_B_r_res}. In this regime, although the range resolution is primarily determined by the signal bandwidth, operating under effective \ac{NF} conditions produces a cleaner range response, i.e., with lower sidelobe levels. Moreover, the larger physical antenna aperture improves the angular resolution. This is evident by comparing the \ac{GLRT}-based radar maps in Fig.~\ref{fig:FF_setup3} and Fig.~\ref{fig:NF_setup3}, when the \ac{ET} is outside the super-resolution region in the simulation setup~$3$ of Fig.~\ref{fig:deltaR_vs_B}.

Fig.~\ref{fig:RMSE_vs_B} shows the \ac{RMSE} of the \ac{ET} centroid as a function of $B$ for the three target setups. In the \ac{FF} case, the behavior isolates the effect of bandwidth alone. For all setups, increasing $B$ (i.e., improving bandwidth-limited range resolution) leads to a marked reduction of the \ac{RMSE}, with additional gains when the \ac{ET} is closer to the transceiver due to a higher \ac{SNR}. In the \ac{NF} case, a similar trend with $B$ is visible in setup~$3$, where the \ac{ET} remains outside the super-resolution region for all considered bandwidths (with the only exception at $B=50\,$MHz, where the \ac{ET} lies close to the boundary of the super-resolution region). Here, the range resolution is dominated by $\Delta r_\mathrm{B}$, and the NF and FF RMSE curves nearly overlap. This confirms that, when the target lies in the effective NF region but outside the super-resolution region, localization accuracy is still essentially governed by the bandwidth-induced range resolution.

Conversely, in setups~$1$ and $2$, where the \ac{ET} lies inside the super-resolution region for almost the entire bandwidth interval, the centroid \ac{RMSE} remains nearly constant with respect to $B$. In this regime, the range resolution is dominated by the \ac{NF} contribution and is essentially independent of the bandwidth. Increasing $B$ therefore does not further improve the resolution, except for $B \lesssim 400\,$MHz, where a larger bandwidth helps suppress sidelobes in the range profile and yields a modest \ac{RMSE} reduction, in line with the discussion in Section~\ref{sec:NF_B_r_res}. Overall, within the super-resolution region, the NF-enhanced range resolution yields a substantial \ac{RMSE} gain over the \ac{FF} baseline, especially for limited bandwidths. Moreover, since the \ac{NF} range resolution is range-dependent, setup~$2$ (with the \ac{ET} closer to the array and higher \ac{SNR}) benefits from finer effective resolution than setup~$1$ (see Fig.~\ref{fig:deltaR_vs_B}), which translates into improved centroid accuracy.

Fig.~\ref{fig:Average_GOSPA_vs_B} shows the average \ac{GOSPA} for all setups in both \ac{FF} and \ac{NF} conditions. As with the \ac{RMSE}, increasing $B$ and placing the \ac{ET} closer to the array generally improves the metric thanks to higher \ac{SNR} and finer range resolution when the target lies in the super-resolution \ac{NF} region. In the \ac{FF} case, the trend mirrors that of Fig.~\ref{fig:RMSE_vs_B} since performance is solely driven by bandwidth-limited resolution. This means that a larger $B$ allows for better discrimination of closely spaced scatterers, reducing both localization error and misdetections.

Unlike the centroid \ac{RMSE}, \ac{NF} operations yield lower \acp{GOSPA} than \ac{FF} across all setups. Notably, even in setup~$3$, where the \ac{ET} lies outside the super-resolution region, \ac{GOSPA} improves significantly under \ac{NF} conditions, although the nominal range resolution remains bandwidth-limited as in \ac{FF}. This improvement is due to additional range–angle focusing beyond pure angle-domain beam steering as well as sidelobe reduction in the range profile (see Section~\ref{sec:NF_B_r_res}), to which \ac{GOSPA} is particularly sensitive due to sidelobe-induced false alarms. As the bandwidth $B$ increases, the higher range resolution reduces \ac{GOSPA} under both \ac{FF} and \ac{NF}, mirroring the \ac{RMSE} trend. However, the decrease is more pronounced in the \ac{NF} setup, as improved resolution combined with \ac{NF}-driven sidelobe mitigation further reduces false alarms.
\section{Conclusions} \label{sec:conclusions}

In this work, we proposed a monostatic wideband \ac{MIMO} \ac{OFDM}-based \ac{ISAC} system at \ac{mmWave} frequencies operating under hybrid near-/far-field conditions: the \ac{Tx} is designed to work in the \ac{FF} with low-complexity beam steering, while the \ac{Rx} operates in the \ac{NF} by means of an \ac{ELAA}. The latter is realized as a \ac{ULA} designed according to an \ac{ELAS} design method, which enlarges the effective aperture by appropriately spacing the receive elements beyond half a wavelength without introducing angle ambiguities. This extends the Fraunhofer distance and enables \ac{NF} sensing gains in spatial resolution and localization accuracy, while remaining compatible with a low-complexity fully digital architecture with only a few \ac{RF} chains. In addition, we analyzed the range resolution of wideband \ac{NF} sensing and identified a super-resolution region where \ac{NF} effects enhance the achievable range resolution beyond the classical bandwidth limit.

Numerical results validate the ability of the \ac{ELAS} design to create effective \ac{NF} conditions with only tens of antenna elements at both ends of the link while naturally avoiding angle ambiguities. Furthermore, we demonstrate that for \ac{ET} modeled as a collection of scatterers, the enhanced NF range resolution within the super-resolution region improves localization accuracy, as measured by the \ac{RMSE} of the estimated target scattering centroid. Finally, although \ac{NF} effects dominate range resolution inside the super-resolution region, waveform bandwidth still plays a key role in shaping the range profile. Moreover, outside this region—where range resolution is bandwidth-limited—the interplay between bandwidth and \ac{NF} effects reduces range sidelobe levels, mitigating sidelobe-induced false alarms, in agreement with the consistently lower \ac{GOSPA} values observed under \ac{NF}.
Overall, the proposed \ac{ELAS}-enabled architecture provides a practical approach for exploiting \ac{NF} super-resolution in wideband sensing with moderate array sizes and fully digital processing.

\ifCLASSOPTIONcaptionsoff
  \newpage
\fi




\balance
\bibliographystyle{IEEEtran}
\bibliography{IEEEabrv,bibliography}

\end{document}